\documentclass[english,man]{apa}
\usepackage[LGR,T1]{fontenc}
\usepackage[latin9]{inputenc}
\usepackage{array}
\usepackage{textcomp}
\usepackage{multirow}
\usepackage{amsmath}
\usepackage{stackrel}
\usepackage{graphicx}
\usepackage{setspace}
\usepackage[authoryear]{natbib}

\makeatletter

\DeclareRobustCommand{\greektext}{%
  \fontencoding{LGR}\selectfont\def\encodingdefault{LGR}}
\DeclareRobustCommand{\textgreek}[1]{\leavevmode{\greektext #1}}
\ProvideTextCommand{\~}{LGR}[1]{\char126#1}

\providecommand{\tabularnewline}{\\}

\helvetica
\author{Author} 
\affiliation{Affiliation} 

\usepackage{kerkis}
\usepackage{apacite}
\usepackage{geometry}
\usepackage{multirow}
\usepackage{rotating}
\usepackage{subfig}
\usepackage{bm}
\usepackage{rotating}
\usepackage{times}
\usepackage{longtable}

\makeatother

\usepackage{babel}
\begin{document}
\title{Distributionally-Weighted Least Squares in Structural Equation Modeling}
\author{Han Du and Peter M. Bentler}
\affiliation{Department of Psychology, University of California, Los Angeles\\ \vspace{1in}Correspondence
should be addressed to Han Du, Pritzker Hall, 502 Portola Plaza, Los
Angeles, CA 90095. Email: hdu@psych.ucla.edu.\\ \vspace{0.5in}This
paper has been presented at the 85th annual Meeting of the Psychometric
Society and the 2020 Meeting of the International Society for Data
Science and Analytics\\ \vspace{0.5in}�American Psychological Association,
2021. This paper is not the copy of record and may not exactly replicate
the authoritative document published in the APA journal. The final
article is available, upon publication, at: https://doi.org/10.1037/met0000388}
\abstract{In real data analysis with structural equation modeling, data are
unlikely to be exactly normally distributed. If we ignore the non-normality
reality, the parameter estimates, standard error estimates, and model
fit statistics from normal theory based methods such as maximum likelihood
(ML) and normal theory based generalized least squares estimation
(GLS) are unreliable. On the other hand, the asymptotically distribution
free (ADF) estimator does not rely on any distribution assumption
but cannot demonstrate its efficiency advantage with small and modest
sample sizes. The methods which adopt misspecified loss functions
including ridge GLS (RGLS) can provide better estimates and inferences
than the normal theory based methods and the ADF estimator in some
cases. We propose a distributionally-weighted least squares (DLS)
estimator, and expect that it can perform better than the existing
generalized least squares, because it combines normal theory based
and ADF based generalized least squares estimation. Computer simulation
results suggest that model-implied covariance based DLS ($DLS_{M}$)
provided relatively accurate and efficient estimates in terms of RMSE.
In addition, the empirical standard errors, the relative biases of
standard error estimates, and the Type I error rates of the Jiang-Yuan
rank adjusted model fit test statistic ($T_{JY}$) in $DLS_{M}$ were
competitive with the classical methods including ML, GLS, and RGLS.
The performance of $DLS_{M}$ depends on its tuning parameter $a$.
We illustrate how to implement $DLS_{M}$ and select the optimal $a$
by a bootstrap procedure in a real data example.}

\maketitle
Structural equation modeling (SEM) is widely used in social and behavioral
research, but its statistical methodology remains marginally capable
of dealing with empirical data encountered in many psychological and
behavioral studies. First, statistics in SEM rely on large sample
size approximation. That is, their use relies on asymptotic properties
as sample size $N$ becomes extremely large ($N\rightarrow\infty$).
However in real data analysis, sample sizes are usually moderate or
even small. Although SEM methods usually provide consistent parameter
estimates and consistent standard error (SE) estimates, the estimates
are not necessarily unbiased with finite sample size. Second, although
data are typically nonnormally distributed (e.g., \citealp{cain2017univariate}),
the mainstream estimators for SEM are still based on normal theory,
such as maximum likelihood (ML) and normal theory based generalized
least squares estimation (GLS). With nonnormally distributed data,
ML and GLS still provide consistent parameter estimates, however their
standard errors and model fit statistics generally are incorrect.
Even when robust standard errors and rescaled-and-adjusted model fit
test statistics are used to correct for non-normality, their performance
in terms of efficiency and Type I error rates with finite sample sizes
has been proved inadequate in a vast literature (e.g., \citealp{jalal2018using,satorra1988scaling,yuan2016structural}).
\footnote{Another type of robust procedure is to weight each observation by
its distance from the center of the data to obtain parameter estimates
and standard errors \citep{yuan1998structural,yuan2012robust}.} 

To relax the normality assumption, \citet{browne1984asymptotically}
proposed an asymptotically distribution free (ADF) estimator for nonnormal
data within the generalized least squares framework. This method
is sometimes called weighted least squares (WLS). The ADF estimator
adopts a completely distribution-free estimate of the asymptotic covariance
matrix of sample covariances (we will expand on this later) and provides
the most asymptotically efficient estimates. However, the efficiency
advantage of the ADF estimator cannot be realized with small and modest
sample sizes. When sample sizes are rather small, there may be serious
convergence problems especially when the number of variables is large,
because the ADF estimator needs to estimate more components (i.e.,
sample fourth-order moments) compared to the methods that rely on
the normality assumption. When sample sizes are modest, the ADF estimator
is unstable because it involves inverting a sample fourth-order moment
matrix. As a consequence, the sampling distribution of ADF estimates
has a large variance and thus standard errors are large compared to
the normal theory based methods. Some researcher have directly used
a (multivariate) t-distribution instead of a (multivariate) normal
distribution to handle data with long tails and tolerate outliers
\citep{song2007maximum,tong2012diagnostics,tong2019robust,zhang2013bayesian}.
In this way, the influence of outliers and distributional deviation
is down-weighted. Rather than using normal distributions, some researchers
used a mixture of distributions which is empirically determined by
the data \citep{lee2006maximum,muthen1999finite}. 

Besides ML, GLS, and WLS, there are other estimators within the generalized
least squares estimation framework, especially, least squares (LS),
diagonally weighted least squares (DWLS), as well as ridge GLS estimation.
We can classify the existing methods into three categories. In the
first category, the methods including ML and GLS rely on a normality
assumption. When the distribution is nonnormal, statistics inference
is not correct. In the second category, WLS makes no distribution
assumption. It enjoys the asymptotic efficiency of ADF but requires
a large sample size to be stable. In the third category, the methods
including LS, DWLS, and ridge GLS use a misspecified loss function.
A misspecified loss function still provides consistent parameter estimates.
In some contexts, such as with ordinal variables, these estimates
can be more accurate and efficient than normal theory based methods
(e.g., \citealp{li2016performance,yuan2019causes}).

We propose a new generalized least squares method, distributionally-weighted
least squares (DLS) estimation. DLS estimation is a combination of
normal theory based and ADF based generalized least squares estimation,
in which the weight matrix of the loss function is the inverse of
a combination of the ADF based and normal theory based estimators
of the covariance matrix of sample covariances ($cov\left(\boldsymbol{s}\right)$).
DLS has three advantages. First, with finite sample sizes, it yields
more efficient estimates than those from the  ADF estimator. Second,
DLS is partially normal theory dependent, which helps to stabilize
the performance of DLS compared to WLS. Third, it balances the information
from the data and the normality assumption. Our context is that of
the typical situation in which sample size substantially exceeds the
number of variables. For the contrary cases, see \citet{deng2018structural}
and \citet{yuan2019causes}.

The outline of this paper is as follows: in the ``Estimators in SEM''
section, an overview is given of some widely used estimators in SEM
and estimators that are related to the current paper. In the ``Distributionally-Weighted
Least Squares'' section, we present the proposed distributionally-weighted
least squares (DLS) estimation. In the ``Model Fit Evaluation''
section, we introduce the model fit statistics that we will explore
in the simulation. In the ``Simulation Study'' section, the performance
of DLS is thoroughly examined via simulations. In the ``Real Data
Example'' section, a real data example is provided to illustrate
the implementation of DLS with bootstrapping in practice. We end the
paper with some concluding remarks in the ``Conclusion'' section.

\section{Estimators in SEM}

We briefly introduce some widely used estimators and some estimators
that are related to our proposed distributionally-weighted least squares
(DLS) method, including maximum likelihood (ML), generalized least
squares (GLS), least squares (LS), weighted least squares (WLS), and
ridge GLS methods. Let $\boldsymbol{x_{1}}$, $\boldsymbol{x_{2}}$,
..., $\boldsymbol{x_{N}}$ be a multivariate random sample of size
$N$ from a $p$-variate population with $E\left(\boldsymbol{x_{i}}\right)=\boldsymbol{\mu}$
and $cov\left(\boldsymbol{x_{i}}\right)=\boldsymbol{\mathbf{\Sigma}}$
for $i=1,...,N$. Let a vector $\boldsymbol{\theta}$ be a $q\times1$
vector containing free parameters in the structural equation model.
The population covariance is assumed to be a function of $\boldsymbol{\theta}$,
therefore $\mathbf{\Sigma}\left(\boldsymbol{\theta}\right)$ is proposed
to fit the data. The population covariance matrix is unknown. In a
sample, we can calculate the model implied covariance matrix $\mathbf{\Sigma}\left(\hat{\boldsymbol{\theta}}\right)$
or the sample covariance matrix $\mathbf{S}$ to estimate the population
covariance matrix $\mathbf{\Sigma}$, 
\begin{equation}
\mathbf{S}=\frac{1}{N}\stackrel[i=1]{N}{\Sigma}\left(\boldsymbol{x_{i}}-\bar{\boldsymbol{x}}\right)\left(\boldsymbol{x_{i}}-\bar{\boldsymbol{x}}\right)',
\end{equation}
where $\bar{\boldsymbol{x}}$ is the sample mean. 

Let $\boldsymbol{s}=vech\left(\mathbf{S}\right)$ be a $p^{*}\times1$
vector with the $p^{*}=p*(p+1)/2$ nonduplicated elements in $\mathbf{S}$
and $\boldsymbol{\sigma}\left(\boldsymbol{\theta}\right)=vech\left[\mathbf{\Sigma}\left(\boldsymbol{\theta}\right)\right]$
be a $p^{*}\times1$ vector with the nonduplicated elements in $\mathbf{\Sigma}\left(\boldsymbol{\theta}\right)$.
By the multivariate central limit theorem, $\boldsymbol{s}$ converges
in distribution to a normal distribution: 
\begin{equation}
\sqrt{N}\left(\boldsymbol{s}-\boldsymbol{\sigma}\left(\boldsymbol{\theta}\right)\right)\overset{D}{\rightarrow}N\left(0,\mathbf{\Gamma}\right),\label{eq: asy}
\end{equation}
where $\mathbf{\Gamma}$ is the $p^{*}\times p^{*}$ asymptotic covariance
matrix of $\sqrt{N}\left(\boldsymbol{s}-\boldsymbol{\sigma}\left(\boldsymbol{\theta}\right)\right)$.
An element of $\mathbf{\Gamma}$ is $\left\{ \mathbf{\Gamma}\right\} _{hj,kl}=cov(\sqrt{N}\left(\mathbf{S}_{hj}-\mathbf{\varSigma}_{hj}\right),\sqrt{N}\left(\mathbf{S}_{kl}-\mathbf{\varSigma}_{kl}\right))=\sigma_{hjkl}-\sigma_{hj}\sigma_{kl}$
with $\sigma_{hjkl}=E\left(\left\{ x_{h}-E\left(x_{h}\right)\right\} \left\{ x_{j}-E\left(x_{j}\right)\right\} \left\{ x_{k}-E\left(x_{k}\right)\right\} \left\{ x_{l}-E\left(x_{l}\right)\right\} \right)$
and $\sigma_{hj}=E\left(\left\{ x_{h}-E\left(x_{h}\right)\right\} \left\{ x_{j}-E\left(x_{j}\right)\right\} \right)=\mathbf{\mathbf{\varSigma}}_{hj}$
\citep{browne1984asymptotically}. In practice, the ADF estimator
uses a consistent estimator of \textgreek{G} with elements 
\begin{equation}
\left\{ \hat{\Gamma}_{ADF}\right\} _{hj,kl}=s_{hjkl}-s_{hj}s_{kl},\label{eq: estimated ADF}
\end{equation}
where $s_{hjkl}=\frac{1}{N}\stackrel[i=1]{N}{\Sigma}\left(\left\{ x_{h,i}-\bar{x}_{h}\right\} \left\{ x_{j,i}-\bar{x}_{j}\right\} \left\{ x_{k,i}-\bar{x}_{k}\right\} \left\{ x_{l,i}-\bar{x}_{l}\right\} \right)$
is the sample fourth moment and $s_{hj}=\frac{1}{N}\stackrel[i=1]{N}{\Sigma}\left(\left\{ x_{h,i}-\bar{x}_{h}\right\} \left\{ x_{j,i}-\bar{x}_{j}\right\} \right)$
is an element of $\mathbf{S}$. 

\begin{table}
\caption{11 methods in the simulation}

\begin{tabular}{ccc}
\hline 
Method & Discrepancy/Loss function  & SE Estimates\tabularnewline
\hline 
\hline 
$DLS_{S}$ & $\left[\boldsymbol{s-\sigma\left(\theta\right)}\right]'\left((1-a)\mathbf{\hat{\Gamma}_{ADF}}+a\mathbf{\hat{\Gamma}_{N.S}}\right)^{-1}\left[\boldsymbol{s-\sigma\left(\theta\right)}\right]$ & \multirow{2}{*}{$a=1$ Standard, $a\neq1$ Sandwich }\tabularnewline
\cline{1-2} \cline{2-2} 
$DLS_{M}$ & $\left[\boldsymbol{s-\sigma\left(\theta\right)}\right]'\left((1-a)\mathbf{\hat{\Gamma}_{ADF}}+a\mathbf{\hat{\Gamma}_{N.M}}\right)^{-1}\left[\boldsymbol{s-\sigma\left(\theta\right)}\right]$ & \tabularnewline
\hline 
$GLS_{S}$ & $\left[\boldsymbol{s-\sigma\left(\theta\right)}\right]'\left(\mathbf{\hat{\Gamma}_{N.S}}\right)^{-1}\left[\boldsymbol{s-\sigma\left(\theta\right)}\right]$ & \multirow{9}{*}{Sandwich }\tabularnewline
\cline{1-2} \cline{2-2} 
$GLS_{M}$ & $\left[\boldsymbol{s-\sigma\left(\theta\right)}\right]'\left(\mathbf{\hat{\Gamma}_{N.M}}\right)^{-1}\left[\boldsymbol{s-\sigma\left(\theta\right)}\right]$ & \tabularnewline
\cline{1-2} \cline{2-2} 
$WLS$ & $\left[\boldsymbol{s-\sigma\left(\theta\right)}\right]'\left(\mathbf{\hat{\Gamma}_{ADF}}\right)^{-1}\left[\boldsymbol{s-\sigma\left(\theta\right)}\right]$ & \tabularnewline
\cline{1-2} \cline{2-2} 
$RGLS_{D}$ & $\left[\boldsymbol{s-\sigma\left(\theta\right)}\right]'\left((1-a)\mathbf{\hat{\Gamma}_{ADF}}+a\times diag(\mathbf{\hat{\Gamma}_{ADF}})\right)^{-1}\left[\boldsymbol{s-\sigma\left(\theta\right)}\right]$ & \tabularnewline
\cline{1-2} \cline{2-2} 
$RGLS_{I}$ & $\left[\boldsymbol{s-\sigma\left(\theta\right)}\right]'\left((1-a)\mathbf{\hat{\Gamma}_{ADF}}+a\mathbf{I}\right)^{-1}\left[\boldsymbol{s-\sigma\left(\theta\right)}\right]$ & \tabularnewline
\cline{1-2} \cline{2-2} 
$LS$ & $\left[\boldsymbol{s-\sigma\left(\theta\right)}\right]'\mathbf{I}\left[\boldsymbol{s-\sigma\left(\theta\right)}\right]$ & \tabularnewline
\cline{1-2} \cline{2-2} 
$ML_{S}$ & \multirow{3}{*}{$tr\left[\mathbf{S\Sigma^{-1}}\left(\boldsymbol{\theta}\right)\right]-log\left|\mathbf{S\Sigma^{-1}}\left(\boldsymbol{\theta}\right)\right|-p$} & \tabularnewline
$ML_{O.M}$ &  & \tabularnewline
$ML_{E.M}$ &  & \tabularnewline
\hline 
\end{tabular}
\end{table}

\subsection{Maximum Likelihood Estimation}

Maximum likelihood (ML) estimation minimizes a function which measures
the discrepancy between $\mathbf{\Sigma}\left(\boldsymbol{\theta}\right)$
 and $\mathbf{S}$ (\citealp{browne1974generalized,browne1984asymptotically,joreskog1967some,joreskog1969general}),
\begin{equation}
F_{ML}\left(\boldsymbol{\theta}\right)=tr\left[\mathbf{S\Sigma^{-1}}\left(\boldsymbol{\theta}\right)\right]-log\left|\mathbf{S\Sigma^{-1}}\left(\boldsymbol{\theta}\right)\right|-p.
\end{equation}
As a variant of ML, Yuan and Chan (2008)\nocite{yuan2008structural}
proposed a ridge ML estimation by replacing $\mathbf{S}$ in the discrepancy
function with $\mathbf{S}_{a}=\mathbf{S}+a\mathbf{I}$, where $a\in\left[0,\infty\right)$
is a ridge tuning parameter and $\mathbf{I}$ is a $p\times p$ identity
matrix. Yuan and Chan (2008)\nocite{yuan2008structural} summarized
that ridge ML estimation does not appropriately handle the variance
of $\mathbf{S}$ because the variance of $\mathbf{S}$ requires the
fourth-order moments information as in Equation (\ref{eq: estimated ADF}).
Both ML and ridge ML use only the sample covariance matrix and do
not involve sample fourth-order moments in parameter estimation. 

Let $\boldsymbol{\dot{\sigma}}\left(\boldsymbol{\theta}\right)$ be
the first-order derivative of $\boldsymbol{\sigma}\left(\boldsymbol{\theta}\right)$
with respect to $\boldsymbol{\theta}$ and be a matrix with dimension
$p^{*}\times q$ where $q$ is the number of free parameters (i.e.,
$\frac{\partial\boldsymbol{\sigma\left(\theta\right)}}{\partial\boldsymbol{\theta'}}$;
Jacobian matrix). The standard error (SE) is the square root of the
diagonal elements of $\left(\boldsymbol{\dot{\sigma}}\left(\boldsymbol{\theta}\right)'\mathbf{I_{\sigma}}\boldsymbol{\dot{\sigma}}\left(\boldsymbol{\theta}\right)\right)^{-1}/N$
where $\mathbf{I_{\sigma}}$ is the information matrix for the structured
model. One can use the expected or observed information matrix to
estimate the standard errors. Additionally, to calculate the expected/observed
information matrix, one can choose to use the sample covariance matrix
$\mathbf{S}$ or the model implied covariance matrix $\mathbf{\Sigma}\left(\hat{\boldsymbol{\theta}}\right)$.
But with the sample covariance matrix $\mathbf{S}$, the expected
and observed information matrices become the same. Hence, there are
three combinations: $ML_{S}$ ($\mathbf{S}$), $ML_{O.M}$ (observed
information and $\mathbf{\Sigma}\left(\hat{\boldsymbol{\theta}}\right)$),
and $ML_{E.M}$ (expected information and $\mathbf{\Sigma}\left(\hat{\boldsymbol{\theta}}\right)$),
which yield different sets of standard error estimates.

Robust statistics, such as sandwich type standard errors \citep{hardin2003sandwich,huber1967behavior,white1980heteroskedasticity,white1982maximum}
and rescaled-and-adjusted model fit test statistics \citep{satorra1986some,satorra1988scaling,satorra1994corrections,jiang2017four,yuan2017empirically},
make use of the sample fourth-order moment information. The sandwich
standard error is obtained from the diagonals of the sandwich covariance
matrix,
\begin{equation}
cov\left(\sqrt{N}\hat{\boldsymbol{\theta}}_{ML}\right)=\left(\boldsymbol{\dot{\sigma}}\left(\boldsymbol{\theta}\right)'\mathbf{I_{\sigma}}\boldsymbol{\dot{\sigma}}\left(\boldsymbol{\theta}\right)\right)^{-1}\boldsymbol{\dot{\sigma}}\left(\boldsymbol{\theta}\right)'\mathbf{I_{\sigma}}\mathbf{\Gamma}_{ADF}\mathbf{I_{\sigma}}\boldsymbol{\dot{\sigma}}\left(\boldsymbol{\theta}\right)\left(\boldsymbol{\dot{\sigma}}\left(\boldsymbol{\theta}\right)'\mathbf{I_{\sigma}}\boldsymbol{\dot{\sigma}}\left(\boldsymbol{\theta}\right)\right)^{-1}.
\end{equation}
We use the estimator $\hat{\mathbf{\Gamma}}_{ADF}$ to replace $\mathbf{\Gamma}_{ADF}$
in practice.

\subsection{Generalized Least Squares Estimation}

A class of generalized least squares (GLS) loss functions \citep{browne1974generalized}
is 
\begin{equation}
F_{GLS}\left(\boldsymbol{\theta}\right)=\left[\boldsymbol{s-\sigma\left(\theta\right)}\right]'\hat{\mathbf{W}}\left[\boldsymbol{s-\sigma\left(\theta\right)}\right],\label{eq:gls loss}
\end{equation}
where $\hat{\mathbf{W}}$ is a weight matrix of size $p^{*}\times p^{*}$
that can take various forms. The best population weight matrix $\mathbf{W}$
is the inverse of the asymptotic covariance matrix of $\sqrt{N}\left(\boldsymbol{s}-\boldsymbol{\sigma}\left(\boldsymbol{\theta}\right)\right)$,
$\mathbf{\Gamma}^{-1}$. The loss function in Equation (\ref{eq:gls loss})
is misspecified whenever $\hat{\mathbf{W}}^{-1}$ is not a consistent
estimate of $\mathbf{\Gamma}$ \citep{savalei2014understanding}.
When $\hat{\mathbf{W}}=\mathbf{I}$, estimation becomes least squares
(LS) estimation. It is very unlikely that the asymptotic covariance
matrix of sample covariances is an identity matrix. Due to the misspecification,
sandwich standard errors usually accompany LS estimates.

The ADF estimator in \citet{browne1984asymptotically} can be used
to specify the weight matrix $\hat{\mathbf{W}}=\hat{\mathbf{\Gamma}}_{ADF}^{-1}$,
as illustrated in Equation (\ref{eq: estimated ADF}). Although the
ADF estimator is known to be asymptotically efficient when $N\rightarrow\infty$
and $p/N\rightarrow0$, with finite sample sizes, it is unstable and
the empirical standard errors can be much greater than those of normal
theory based estimators even when data are nonnormally distributed
\citep{yang2019optimizing,yuan1997improving,yuan2016structural}.
We refer to this method as the weighted least squares estimation (WLS;
\citealp{rosseel2012lavaan}). 

Under the normal theory assumption, $s$ has an asymptotic covariance
matrix that has a relatively simple form, that is the asymptotic covariance
matrix is $\left\{ \mathbf{\Gamma}_{N}\right\} _{hj,kl}=\sigma_{hk}\sigma_{jl}+\sigma_{hl}\sigma_{jk}$.
Hence, with the normal theory assumption, $\hat{\mathbf{W}}$ is specified
to be $\hat{\mathbf{\Gamma}}_{N}^{-1}$. In samples, $\hat{\mathbf{\Gamma}}_{N}$
can be either estimated by sample covariances $\left\{ \hat{\mathbf{\Gamma}}_{N.S}\right\} _{hj,kl}=s_{hk}s_{jl}+s_{hl}s_{jk}$
or estimated by the model implied covariances $\left\{ \hat{\mathbf{\Gamma}}_{N.M}\right\} _{hj,kl}=\hat{\sigma}_{hk}\hat{\sigma}_{jl}+\hat{\sigma}_{hl}\hat{\sigma}_{jk}$.
We refer to the estimation using sample covariances as $GLS_{S}$
and the estimation using model implied covariances as $GLS_{M}$.

The loss functions of the aforementioned GLS methods plus some to
be explained below are summarized in Table 1. The sandwich standard
error is obtained from the square root of the diagonals of the sandwich
covariance matrix,
\begin{equation}
cov\left(\sqrt{N}\hat{\boldsymbol{\theta}}_{GLS}\right)=\left(\boldsymbol{\dot{\sigma}}\left(\boldsymbol{\theta}\right)'\mathbf{W}\boldsymbol{\dot{\sigma}}\left(\boldsymbol{\theta}\right)\right)^{-1}\boldsymbol{\dot{\sigma}}\left(\boldsymbol{\theta}\right)'\mathbf{W}\mathbf{\Gamma}_{ADF}\mathbf{W}\boldsymbol{\dot{\sigma}}\left(\boldsymbol{\theta}\right)\left(\boldsymbol{\dot{\sigma}}\left(\boldsymbol{\theta}\right)'\mathbf{W}\boldsymbol{\dot{\sigma}}\left(\boldsymbol{\theta}\right)\right)^{-1}.\label{eq:GLS se}
\end{equation}
We use $\hat{\mathbf{\Gamma}}_{ADF}$ to replace $\mathbf{\Gamma}_{ADF}$
and $\hat{\mathbf{W}}$ to replace $\mathbf{W}$ in practice. For
WLS, Equation (\ref{eq:GLS se}) simplifies to $cov\left(\sqrt{N}\hat{\boldsymbol{\theta}}_{GLS}\right)=\left(\boldsymbol{\dot{\sigma}}\left(\boldsymbol{\theta}\right)'\hat{\mathbf{W}}_{ADF}\boldsymbol{\dot{\sigma}}\left(\boldsymbol{\theta}\right)\right)^{-1}$.

\subsection{Ridge Generalized Least Squares Estimation}

When the population distribution is unknown, the assumption of normality
is unlikely to be supported. However, using a distribution free estimator
leads to unstable performance. Yuan and Chan (2016)\nocite{yuan2008structural}
 and Yuan, Jiang, and Cheng (2017)\nocite{yuan2017more} proposed
two types of ridge GLS method (RGLS) for continuous and ordinal variables
to stabilize the performance of the ADF estimator. To increase the
efficiency of the ADF estimator (i.e., the WLS method), Yuan and Chan
(2016)\nocite{yuan2016structural} and Yuan, Jiang, and Cheng (2017)\nocite{yuan2017more}
added components to the diagonals of $\hat{\mathbf{W}}$. More specifically,
$\hat{\mathbf{W}}$ is constructed as $\left((1-a)\hat{\mathbf{\Gamma}}_{ADF}+a\mathbf{I}\right)^{-1}$
or $\left((1-a)\mathbf{\hat{\Gamma}}_{ADF}+a\times diag(\hat{\mathbf{\Gamma}}_{ADF})\right)^{-1}$
where $a\in\left[0,1\right]$ is a ridge tuning parameter. The former
one is referred to as $RGLS_{I}$ and the latter one is referred to
as $RGLS_{D}$. RGLS gains the stability of employing simple weight
matrices (i.e., $\mathbf{I}$ and $diag(\hat{\mathbf{\Gamma}}_{ADF})$)
and the asymptotic efficiency of the ADF estimator. For both $RGLS_{I}$
and $RGLS_{D}$, the empirical performance depends on the value of
the ridge tuning parameter, although the estimates remain consistent.
When $a=1$ in $RGLS_{I}$, $RGLS_{I}$ becomes $LS$. When $a=0$
in $RGLS_{I}$ or $RGLS_{D}$, $RGLS_{I}$ or $RGLS_{D}$ becomes
$WLS$. Yang and Yuan (2019)\nocite{yang2019optimizing}, Yuan and
Chan (2016)\nocite{yuan2016structural}, and Yuan, Jiang, and Cheng
(2017)\nocite{yuan2017more} suggest that one can select the optimal
$a$ based on the efficiency and accuracy of parameter estimates.
The $a$ with the minimum root mean square error (RMSE) is the optimal
$a$ (denoted as $a_{s}$). RMSE considers both efficiency and accuracy
in estimation. The reason for considering both efficiency and accuracy
is that estimates are not necessarily unbiased with a small or moderate
sample size, and the variance of SE estimates is not a good index
for biased estimates.

Yang and Yuan (2019)\nocite{yang2019optimizing}, Yuan and Chan (2016)\nocite{yuan2008structural}
, and Yuan, Jiang, and Cheng (2017)\nocite{yuan2017more} found that
$a_{s}$ depend on all aspects of the data and model, including the
number of variables, the number of factors, and the population distribution.
In practice, $a_{s}$ is unknown and needs to be estimated. Currently,
there are two ways to estimate $a_{s}$. First, one can use the bootstrap
procedure to create multiple samples and calculate empirical RMSE
to select $a_{s}$ \citep{yuan2016structural,yuan2017more}. Second,
one can obtain a mapping function between $a_{s}$ and all data/model
features by an extensive simulation. In a real data analysis, such
a mapping function can be adopted to estimate $a_{s}$ based on the
data/model information in that real data set \citep{jiang2018ridge,yang2018ridge,yang2019optimizing}. 

Simulations from Yuan and Chan (2016)\nocite{yuan2016structural}
showed that with $a_{s}$, $RGLS_{I}$ performed better than $RGLS_{D}$,
$LS$, and $WLS$ in terms of the efficiency and accuracy of parameter
estimates. The sandwich SEs from $RGLS_{I}$ were close to the empirical
SEs across replications. Additionally, the convergence rates of $RGLS_{I}$
and $RGLS_{D}$ were higher than the ones of $WLS$.

\section{Distributionally-Weighted Least Squares}

We introduce the procedure of distributionally-weighted least squares
(DLS) in this section. The DLS method falls in the generalized least
squares estimation framework. Similar to the RGLS methods, DLS balances
performance under finite sample size with asymptotic performance.
The research on RGLS by Yuan and Chan (2016)\nocite{yuan2008structural}
 and Yuan, Jiang, and Cheng (2017)\nocite{yuan2017more} sheds light
on the current study. $\mathbf{\hat{\Gamma}_{ADF}}$ has large variability
and cannot demonstrate its merits with small to moderate sample sizes.
By adding information from normal theory based estimated $\mathbf{\Gamma}$
($\mathbf{\hat{\Gamma}_{N}}$), we can stabilize the performance from
the ADF estimator and improve efficiency with small to moderate sample
sizes. Using the loss function of general GLS in Equation (\ref{eq:gls loss}),
we propose to specify the weight function as 
\begin{equation}
\hat{\mathbf{W}}=\left((1-a)\mathbf{\hat{\Gamma}_{ADF}}+a\mathbf{\hat{\Gamma}_{N}}\right)^{-1},\label{eq:weight in DLS}
\end{equation}
where $a$ is a tuning parameter. DLS provides consistent estimates
of $\boldsymbol{\theta}$ regardless of the value of $a$, and DLS
is more efficient (i.e., smaller standard errors) than WLS with $\hat{\mathbf{W}}=\hat{\mathbf{\Gamma}}_{ADF}$
unless $a=0$. Although $\hat{\boldsymbol{\theta}}$ needs to be solved
iteratively, $\hat{\boldsymbol{\theta}}$ is asymptotically equivalent
to $\boldsymbol{\left(\hat{\dot{\sigma}}'\hat{\mathbf{W}}\hat{\dot{\sigma}}\right)^{-1}\hat{\dot{\sigma}}'\hat{\mathbf{W}}\left(s-\hat{\sigma}\right)}$
with $\mathbf{\dot{\sigma}\left(\theta\right)}=\frac{\partial\boldsymbol{\sigma\left(\theta\right)}}{\partial\boldsymbol{\theta'}}$.
As mentioned above, $\hat{\mathbf{\Gamma}}_{N}$ can be estimated
by sample covariances ($\left\{ \hat{\mathbf{\Gamma}}_{N.S}\right\} _{ij,kl}=s_{ik}s_{jl}+s_{il}s_{jk}$
based on $\mathbf{S}$) or the model implied covariances ($\left\{ \hat{\mathbf{\Gamma}}_{N.M}\right\} _{ij,kl}=\hat{\sigma}_{ik}\hat{\sigma}_{jl}+\hat{\sigma}_{il}\hat{\sigma}_{jk}$
based on $\mathbf{\Sigma}\left(\hat{\boldsymbol{\theta}}\right)$).
Thus, there are sample covariance based DLS ($DLS_{S}$) and model-implied
covariance based DLS ($DLS_{M}$) depending on how we calculate $\hat{\mathbf{\Gamma}}_{N}$.
Given an $a$, we can use a Newton method to minimize the loss function
and estimate the parameters. When $a$ is not 1, the SE estimate is
calculated as the sandwich standard error based on Equation (\ref{eq:GLS se}).
When $a$ is 1, $\mathbf{\Gamma_{N}}$ is selected and the normal
assumption is used in the weight function, therefore we use the standard
non-robust SE estimate.

The empirical performance of DLS depends on $a$. A larger $a$ provides
finite sample stability, whereas a smaller $a$ provides asymptotic
efficiency. When $a=1$ in DLS, $\mathbf{\hat{W}}$ in Equation (\ref{eq:weight in DLS})
simplifies to be $\mathbf{\hat{\Gamma}_{N}^{-1}}$ and leads to normal
theory based generalized least squares estimation. More specifically,
$DLS_{S}$ simplifies to the sample covariance based GLS ($GLS_{S}$)
and $DLS_{M}$ simplifies to the model-implied covariance based GLS
($GLS_{M}$). When $a=0$ in DLS, $\mathbf{\hat{W}}$ simplifies to
be $\mathbf{\hat{\Gamma}_{ADF}^{-1}}$ and leads to weighted least
squares (WLS). Because DLS combines the strengths of the normal theory
based GLS and distribution free (i.e., ADF) based GLS, we expect it
to yield more efficient and accurate parameter estimates than ML,
WLS, normal theory based GLS ($GLS_{M}$ and $GLS_{S}$), and LS when
data are nonnormal, and to yield similar estimates as ML and normal
theory based GLS when data are normal. 

Following Yuan and Chan (2016)\nocite{yuan2008structural}, Yuan,
Jiang, and Cheng (2017)\nocite{yuan2017more}, and Yang and Yuan (2019)\nocite{yang2019optimizing},
we select the optimal $a$ ($a_{s}$) as the one corresponding to
the most efficient and accurate parameter estimates. We can quantify
both the efficiency and accuracy of parameter estimates by the root
mean square error (RMSE). When data are normal, we expect that $a_{s}$
is 1 or close to 1, because the information in $\mathbf{\hat{\Gamma}_{ADF}^{-1}}$
should not be able to improve the efficiency and accuracy of the parameter
estimates. When data are nonnormal, $\hat{\mathbf{\Gamma}}_{N}$ is
misspecified, therefore we expect that $a_{s}$ is not 1 and $\mathbf{\hat{\Gamma}_{ADF}^{-1}}$
steps in to provide more efficient and accurate parameter estimates. 

\section{Model Fit Evaluation}

In SEM, researchers usually evaluate whether the model fits the data
well. The standard model fit statistic under the normality assumption
is $T=\left(N-1\right)F\left(\hat{\boldsymbol{\theta}}\right)$, where
$F\left(\hat{\boldsymbol{\theta}}\right)$ is the discrepancy ($F_{ML}\left(\boldsymbol{\theta}\right)$)
or loss function ($F_{GLS}\left(\boldsymbol{\theta}\right)$). $T$
asymptotically follows $\chi_{df}^{2}$ where $df=p^{*}-q$. When
the distribution of data is not normal, the asymptotic distribution
of $T$ is a weighted sum of $df$ independent $\chi_{1}^{2}$. Let
$\mathbf{U=W-W\dot{\sigma}\left(\theta\right)\left(\dot{\sigma}\left(\theta\right)'W\dot{\sigma}\left(\theta\right)\right)^{-1}\dot{\sigma}\left(\theta\right)'W}$.
$\mathbf{W}$ is the population counter part of the $\mathbf{\hat{W}}$
in the loss function for the GLS related methods, or the population
counter part of the information matrix $\mathbf{I_{\sigma}}$ for
the ML methods. The mean of the asymptotic distribution is $tr\left(\mathbf{U\Gamma}\right)$
\citep{satorra1988scaling}. The key idea of the rescaled-and-adjusted
test statistics is to adjust test statistics so that we can more closely
approximate a reference $\chi^{2}$ distribution. The general form
is $T_{R}=T/c$ with $c$ as the adjustment constant. There are several
widely used rescaled-and-adjusted test statistics in the literature.
The most widely used one is the Satorra--Bentler statistic $T_{SB}$
\citep{satorra1986some,satorra1988scaling,satorra1994corrections}
with
\begin{equation}
c_{SB}=tr\left(\mathbf{\hat{U}\hat{\Gamma}_{ADF}}\right)/df\label{eq:c_sb}
\end{equation}
where $\hat{\mathbf{U}}$ and $\hat{\mathbf{\Gamma}}_{ADF}$ are consistent
estimates of $\mathbf{U}$ and $\mathbf{\Gamma}$, respectively. $c_{SB}$
rescales the asymptotic distribution to have a mean of $df$. $T_{SB}=T/c_{SB}$
is referred to $\chi_{df}^{2}$. Satorra and Bentler (1988)\nocite{satorra1988scaling}
proposed another corrected statistic that has both the mean and variance
of the test statistic adjusted, $T_{MVA}=T/c_{MVA}$ with
\begin{align}
c_{MVA} & =tr\left(\left(\mathbf{\hat{U}\hat{\Gamma}_{ADF}}\right)^{2}\right)/tr\left(\mathbf{\hat{U}\hat{\Gamma}_{ADF}}\right).\label{eq:c_mva}
\end{align}
$T_{MVA}$ is referred to $\chi_{df^{*}}^{2}$ where $df^{*}=\left[tr\left(\mathbf{\hat{U}\hat{\Gamma}_{ADF}}\right)\right]^{2}/tr\left(\left(\mathbf{\hat{U}\hat{\Gamma}_{ADF}}\right)^{2}\right)$.
In practice, $\mathbf{\hat{U}\hat{\Gamma}_{ADF}}$ can be rank-deficient.
If the rank of $\mathbf{\hat{U}\hat{\Gamma}_{ADF}}$ is smaller than
$df$, Equations (\ref{eq:c_sb}) and (\ref{eq:c_mva}) are not valid.
Hence, Jiang and Yuan (2017)\nocite{jiang2017four} proposed to estimate
the average eigenvalues of $\mathbf{U\Gamma}$ by replacing $df$
with $rank\left(\mathbf{\hat{U}\hat{\Gamma}_{ADF}}\right)$, $T_{JY}=T/c_{JY}$
with
\begin{align}
c_{JY} & =tr\left(\mathbf{\hat{U}\hat{\Gamma}_{ADF}}\right)/rank\left(\mathbf{\hat{U}\hat{\Gamma}_{ADF}}\right).\label{eq:JY}
\end{align}
$T_{JY}$ is compared to $\chi_{df}^{2}$ in Jiang and Yuan (2017)\nocite{jiang2017four}.
However, we propose to compare $T_{JY}$ to $\chi_{rank\left(\mathbf{\hat{U}\hat{\Gamma}_{ADF}}\right)}^{2}$
because $T_{JY}$ is re-centered to have a mean of $rank\left(\mathbf{\hat{U}\hat{\Gamma}_{ADF}}\right)$.
In the simulation, we will explore the performance of $T_{JY}$ compared
to both $\chi_{df}^{2}$ and $\chi_{rank\left(\mathbf{\hat{U}\hat{\Gamma}_{ADF}}\right)}^{2}$.

\section{Simulation Study}

\subsection{Simulation Design}

This simulation study examines the performance of the proposed distributionally-weighted
least squares estimation, $DLS_{M}$ and $DLS_{S}$. We varied the
values of the following four factors: the total number of variables
($p=5$, 15, and 30), the number of factors ($m=1$ and 3), the sample
size ($N$ ranging from 40 to 1000), and the distributional conditions
(a normal distribution, an elliptical distribution, and two skewed
distributions due to either skewed factors or skewed errors). Specifically,
in a model with a larger $p$, the sample size $N$ needs to be greater
in order to obtain a set of converged parameter estimates, therefore
the conditions of $N$ were nested within $p$. In total, we considered
18 conditions of $N$, $p$, and $m$ (Table 2 illustrates all the
conditions). $x$ was simulated from a confirmatory factor analysis
(CFA) model
\begin{equation}
\boldsymbol{x}=\boldsymbol{\mu}+\mathbf{\Lambda}\boldsymbol{\xi}+\boldsymbol{\varepsilon},\label{eq:CFA model}
\end{equation}
where $\boldsymbol{\mu}$ is a $p\times1$ vector of means (fixed
at 0 when generating the data and not estimated), $\mathbf{\Lambda}$
is a $p\times m$ vector of factor loadings, $\boldsymbol{\xi}$ is
a $m\times1$ vector of factor scores, and $\boldsymbol{\varepsilon}$
is a $p\times1$ vector of independent measurement errors for $p$
variables. Let $\mathbf{\Phi}=cov\left(\boldsymbol{\xi}\right)$ and
$\mathbf{\Psi}=cov\left(\boldsymbol{\varepsilon}\right)$, then the
corresponding population covariance matrix of $x$ is 
\begin{equation}
\mathbf{\Sigma}=\mathbf{\mathbf{\Lambda}\mathbf{\Phi}\mathbf{\Lambda}}'+\mathbf{\Psi}.\label{eq:covariance equation}
\end{equation}
We specified a simple cluster structure, with each factor having the
same number of free non-zero loadings. For example, when $p=15$ and
$m=3$, each factor had 5 nonzero loadings. Following Yang and Yuan
(2019)\nocite{yang2019optimizing}, the population values of the factor
loadings were randomly sampled from .70 to .95, with an interval of
.05. The correlations/covariances between the factors were specified
as 0.50 and the variances of the factors were specified as 1. $\mathbf{\Psi}$
was calculated to ensure that the diagonal elements of $\mathbf{\mathbf{\mathbf{\Sigma}}}$
were 1. Following Yuan and Chan (2016)\nocite{yuan2016structural}
and Yang and Yuan (2019)\nocite{yang2019optimizing}, when the distributional
condition was normal, $\xi=\mathbf{\Phi}^{1/2}Z_{\xi}$ and $\varepsilon=\mathbf{\Psi}^{1/2}Z_{\varepsilon}$
where $\mathbf{\Phi}^{1/2}\mathbf{\Phi}^{1/2}=\mathbf{\Phi}$, $\mathbf{\Psi}^{1/2}\mathbf{\Psi}^{1/2}=\mathbf{\Psi}$,
and both $Z_{\xi}$ and $Z_{\varepsilon}$ followed a standard normal
distribution $N(0,1)$. When the distributional condition was elliptical
(symmetric distributions with heavy tails), $\xi=r\mathbf{\Phi}^{1/2}Z_{\xi}$
and $\varepsilon=r\mathbf{\Psi}^{1/2}Z_{\varepsilon}$ with $r\sim\left(3/\chi_{5}^{2}\right)^{1/2}$.
Because $E\left(r^{2}\right)=1$, $E\left(Z_{\xi}\right)=0$, and
$E\left(Z_{\varepsilon}\right)=0$, Equation (\ref{eq:covariance equation})
is still applicable. When the factors were skewed, we also considered
the heavy-tail feature in generating data. In other words, we used
$r\sim\left(3/\chi_{5}^{2}\right)^{1/2}$ to add heavy tails to the
skewed distribution: $\xi=r\mathbf{\Phi}^{1/2}Z_{\xi}$ and $\varepsilon=r\mathbf{\Psi}^{1/2}Z_{\varepsilon}$
where $Z_{\xi}\sim standardized\left(\chi_{1}^{2}\right)$ and $Z_{\varepsilon}\sim N(0,1)$.
When the error were skewed, $\xi=r\mathbf{\Phi}^{1/2}Z_{\xi}$ and
$\varepsilon=r\mathbf{\Psi}^{1/2}Z_{\varepsilon}$ where $Z_{\xi}\sim N(0,1)$
and $Z_{\varepsilon}\sim standardized\left(\chi_{1}^{2}\right)$.
For each condition, we simulated 1000 datasets. We conducted the simulation
with R (version 3.6.1). The R code of DLS is provided on (https://github.com/hduquant/lab\_code\_collection/blob/master/DLS/DLS\_code.R)
and the ML estimation is implemented by an R package, \textit{lavaan
(version 0.6-5)} \citep{rosseel2012lavaan}. DLS will soon be available
in the forthcoming version of \textit{lavaan}.

\begin{table}
\caption{Simulation Conditions of $N$, $p$, and $m$ }

\begin{tabular}{ccccccc}
\hline 
\multicolumn{5}{c}{18 Conditions of $N$, $p$, and $m$} &  & \tabularnewline
\hline 
$m$ & $p$ & \multicolumn{3}{c}{$N$} &  & \tabularnewline
\hline 
1 & 5 & \multicolumn{3}{c}{40, 60, 100, 200, 300, 500, 1000} &  & \tabularnewline
3 & 15 & \multicolumn{3}{c}{40, 60, 100, 200, 300, 500, 1000} &  & \tabularnewline
3 & 30 & \multicolumn{3}{c}{100, 300, 500, 1000} &  & \tabularnewline
\hline 
\hline 
\multicolumn{5}{c}{4 Distributional Conditions } &  & \tabularnewline
\hline 
\multicolumn{2}{c}{} & $Z_{\xi}$ & $Z_{\varepsilon}$ & $r$ & Skewness  & Kurtosis\tabularnewline
\hline 
\multicolumn{2}{c}{Normal} & $N(0,1)$ & $N(0,1)$ & - & 0.995  & 0.999\tabularnewline
\multicolumn{2}{c}{Elliptical} & $N(0,1)$ & $N(0,1)$ & $\left(3/\chi_{df}^{2}\right)^{1/2}$ & 21.947  & 2.270\tabularnewline
\multicolumn{2}{c}{Skewed Factor} & $standardized\left(\chi_{1}^{2}\right)$ & $N(0,1)$ & $\left(3/\chi_{df}^{2}\right)^{1/2}$ & 76.535  & 2.749\tabularnewline
\multicolumn{2}{c}{Skewed Error} & $N(0,1)$ & $standardized\left(\chi_{1}^{2}\right)$ & $\left(3/\chi_{df}^{2}\right)^{1/2}$ & 208.130 &  4.068\tabularnewline
\hline 
\end{tabular}

Note: The average skewness and kurtosis are calculated from the simplest
model ($m=1$ and $p=5$) and the largest sample size ($N=1000$),
based on 1000 replications. Following \citet{yuan2017empirically},
the multivariate skewness is calculated as $\frac{1}{Np(p+1)(p+2)}\stackrel[i=1]{N}{\sum}\stackrel[j=1]{N}{\sum}\left[\left(\mathbf{x_{i}}-\bar{\mathbf{x}}\right)'\mathbf{S}^{-1}\left(\mathbf{x_{j}}-\bar{\mathbf{x}}\right)\right]^{3}$
and the multivariate kurtosis is calculated as $\frac{1}{Np(p+2)}\stackrel[i=1]{N}{\sum}\left[\left(\mathbf{x_{i}}-\bar{\mathbf{x}}\right)'\mathbf{S}^{-1}\left(\mathbf{x_{i}}-\bar{\mathbf{x}}\right)\right]^{2}$.
The population multivariate skewness and kurtosis should be 1 in the
normal case.
\end{table}

We considered the values for the tuning parameter $a$ from 0 to 1
with equal interval of 0.01 (i.e., 0, 0.01,..., 0.99, 1). Hence, there
were 101 $a$ values adopted for each simulated dataset. In estimating
the CFA model in Equation (\ref{eq:CFA model}), all the diagonal
elements of $\mathbf{\Phi}$ were fixed at 1 and all non-zero factor
loadings were freely estimated. Therefore, the $q\times1$ vector
$\boldsymbol{\theta}$ contains all the free parameters in the SEM
model: $p$ free factor loadings, $m\left(m-1\right)/2$ factor covariances,
and $p$ error variances ($q=2p+m\left(m-1\right)/2$). 

There were 11 methods considered in the simulation (see Table 1).
The distributionally-weighted least squares (DLS) estimation can rely
on either the sample covariances ($\mathbf{S}$) or the model-implied
covariances ($\mathbf{\Sigma}\left(\hat{\boldsymbol{\theta}}\right)$)
to obtain the estimated normal theory based asymptotic covariance
matrix of $s$ ($\mathbf{\hat{\Gamma}_{N}}$), therefore there were
sample covariance based DLS ($DLS_{S}$) and model-implied covariance
based DLS ($DLS_{M}$). When $a=1$ in DLS, the sample covariance
and normal theory based GLS ($GLS_{S}$) and model-implied covariance
and normal theory based GLS ($GLS_{M}$) become special cases of DLS.
When $a=0$ in DLS, weighted least squares estimation (WLS) becomes
a special case of DLS. We also considered $RGLS_{D}$ and $RGLS_{I}$
from Yuan and Chan (2016)\nocite{yuan2016structural}. In $RGLS_{I}$,
when $a=1$, $\mathbf{\hat{W}}$ simplifies to be $I$ and leads to
least squares estimation (LS). In addition, we considered normal theory
based maximum likelihood (ML) estimation. Among ML procedures, the
combination of different information matrix (observed or expected
information) and different covariance matrix (sample or model-implied
covariance) yields three methods ($ML_{O.M}$, $ML_{S}$, and $ML_{E.M}$)
with different sets of standard error (SE) estimates but the same
parameter estimates. 

To investigate both the efficiency and accuracy of parameter estimates,
the root mean square error (RMSE) is a widely used index (e.g., \citealp{yuan2016structural,yuan2017empirically,yang2019optimizing}).
Let $\hat{\theta}_{ij}$ be the estimate of the $i$th parameter in
the $j$th replication. The RMSE for each condition was averaged over
all parameters,
\begin{equation}
RMSE=\frac{1}{q}\stackrel[i=1]{i=q}{\sum}\left(\frac{1}{1000}\stackrel[j=1]{j=1000}{\sum}\left(\hat{\theta}_{ij}-\theta_{i}\right)^{2}\right)^{1/2},\label{eq:rmse}
\end{equation}
where $\theta_{i}$ was the true value for the $i$th parameter.

To investigate the performance of the SE estimates of different methods,
we calculated the relative biases of the SE estimates. The true SE
is unknown, therefore we calculated the standard deviation for each
parameter estimate across 1000 replications as the empirical SE of
each parameter. Let $\hat{SE}_{ij}$ be the SE estimate of $i$th
parameter in the $j$th replication and $SE_{i}$ be the empirical
SE of $i$th parameter. The relative biases of the SE estimates were
averaged over all parameters that we are interested in,
\[
Relative\:Bias=\frac{1}{q}\stackrel[i=1]{i=q}{\sum}\left(\left|\frac{\left(\frac{1}{1000}\stackrel[j=1]{j=1000}{\sum}\hat{SE}_{ij}\right)-SE_{i}}{SE_{i}}\right|\right).
\]

For $RGLS_{D}$, $RGLS_{I}$, $LS$, $WLS$, $GLS_{M}$, $GLS_{S}$,
$ML_{O.M}$, $ML_{E.M}$, and $ML_{S}$, the SE estimates were the
sandwich SE estimates. For $DLS_{M}$ and $DLS_{S}$, the SE estimates
were the sandwich SE estimates when $a$ was not 1. When $a$ was
1, the standard SE estimates were adopted. In terms of model fit evaluation,
we considered 4 model fit statistics: the standard model fit statistic
($T$), the Satorra--Bentler test statistic ($T_{SB}$), the mean
and variance adjusted test statistic ($T_{MVA}$), and the Jiang-Yuan
rank adjusted test statistic ($T_{JY}$). 

\subsection{Structure of Simulation Results}

In the results sections, we first summarize convergence issues of
the 11 methods. Second, we present the influence of $a$ on the performance
of $DLS$. Third, we compared RMSEs of parameter estimates, empirical
SEs, biases of SE estimates, and Type I error rates of model fit statistics
across methods, respectively. In the end, we provide our conclusions
from the simulation results. We present all the detailed results in
the supplemental material.

\subsection{Convergence Issue}

The convergence rates of $LS$, $GLS_{M}$, $ML_{O.M}$, $ML_{E.M}$,
$ML_{S}$, $RGLS_{D}$ (with the optimal $a$) and $RGLS_{I}$ (with
the optimal $a$) were all almost 1 across conditions (i.e., > 0.98).
When the sample size $N$ was too small relative to the model complexity
(e.g., $N=40$, $p=15$, $m=3$), $GLS_{S}$ and $DLS_{S}$ could
have convergence rates lower than 0.8 but higher than 0.7, and $WLS$
has no converged results at all because only the ADF estimator was
used. $DLS_{M}$ with the optimal $a$ always had a convergence rate
near 1; with a larger sample size, $DLS_{S}$ went above 0.9. The
influence of different $a$ values on the convergence of $DLS$ will
be expanded upon in next section. We  kept only the converged solutions
among the 1000 replications.

\subsection{Influence of $a$ on $DLS$}

The value of $a$ determines the performance of $DLS_{M}$ and $DLS_{S}$.
We discuss the performance in terms of convergence rates, RMSE, relative
biases of SE, and Type I error rates of test statistics. 

In terms of root mean square errors (RMSE) of parameter estimates,
the optimal $a$ ($a_{s}$) which minimized the average RMSE across
all parameters depended on the distribution and the complexity of
the model. Within each model, $N$ did not obviously change the trajectories
of RMSE along with $a$. $a_{s}$ was consistent between $DLS_{M}$
and $DLS_{S}$, and the minimal RMSE was smaller in $DLS_{M}$. When
the distribution was normal, $a=1$ (or almost 1) in both $DLS_{M}$
and $DLS_{S}$ provided the smallest RMSE, which indicated that the
algorithm heavily weighted on $\hat{\mathbf{\Gamma}}_{N}$. We illustrate
the plots of RMSE of all the parameter estimates for $DLS_{M}$, $DLS_{S}$,
$RGLS_{I}$, and $RGLS_{D}$, when $N=300$, $p=5$ and $m=1$ in
Figure 1 and $N=300$, $p=30$ and $m=3$ in Figure 2 respectively,
as examples of simple and complex models with a moderate sample size.
When the distribution was nonnormal, $a_{s}$ depended on the complexity
of the model. With a simpler model, $a_{s}$ was smaller in $DLS_{M}$
and $DLS_{S}$ (e.g., can be about 0.7; see Figure 1). In a more complex
model, $a_{s}$ was close to 1 in $DLS_{M}$ and $DLS_{S}$ (see Figure
2). For $RGLS_{I}$ and $RGLS_{D}$, $a_{s}$ was smaller in a simpler
model. This indicated that $RGLS$ employed simple weight matrices
(i.e., $\mathbf{I}$ and $diag(\hat{\mathbf{\Gamma}}_{ADF})$) more
strongly in a simpler model. Compared to $RGLS$, $DLS_{M}$ and $DLS_{S}$
were more sensitive to the selection of $a$. 

\begin{figure}
{\footnotesize{}\caption{{\footnotesize{}Plot of root mean square error (RMSE) of model parameter
estimates depending on $a$ when $N=300$, $p=5$ and $m=1$}}
}{\footnotesize\par}

\begin{singlespace}
{\footnotesize{}Figure 1: Plot of root mean square error (RMSE) of
model parameter estimates depending on $a$ when $N=300$, $p=5$
and $m=1$}{\footnotesize\par}
\end{singlespace}

\includegraphics[scale=0.5]{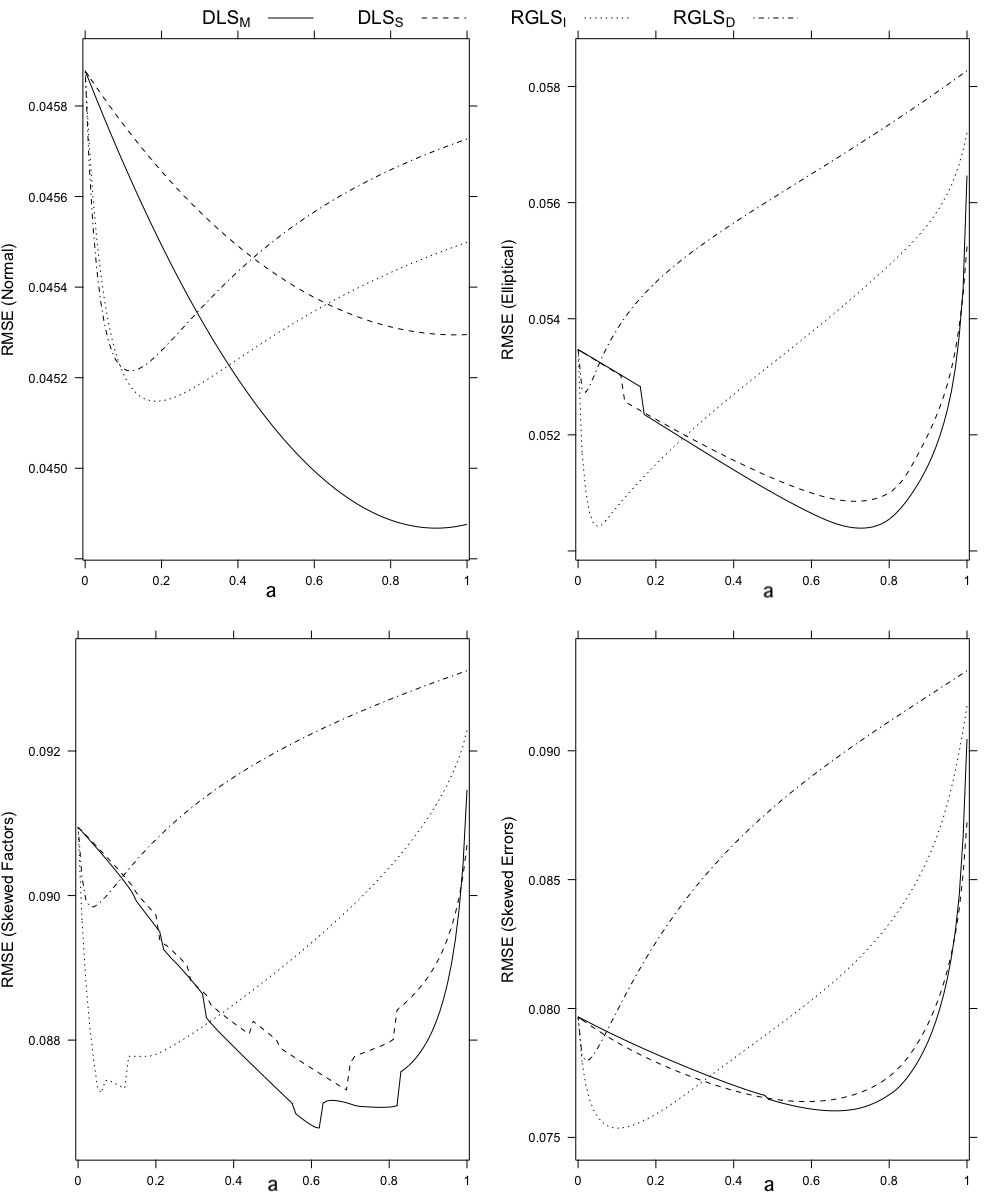}
\end{figure}
\begin{figure}
{\footnotesize{}\caption{{\footnotesize{}Plot of root mean square error (RMSE) of model parameter
estimates depending on $a$ when $N=300$, $p=30$ and $m=3$}}
}{\footnotesize\par}

\begin{singlespace}
{\footnotesize{}Figure 2: Plot of root mean square error (RMSE) of
model parameter estimates depending on $a$ when $N=300$, $p=30$
and $m=3$}{\footnotesize\par}
\end{singlespace}

\includegraphics[scale=0.5]{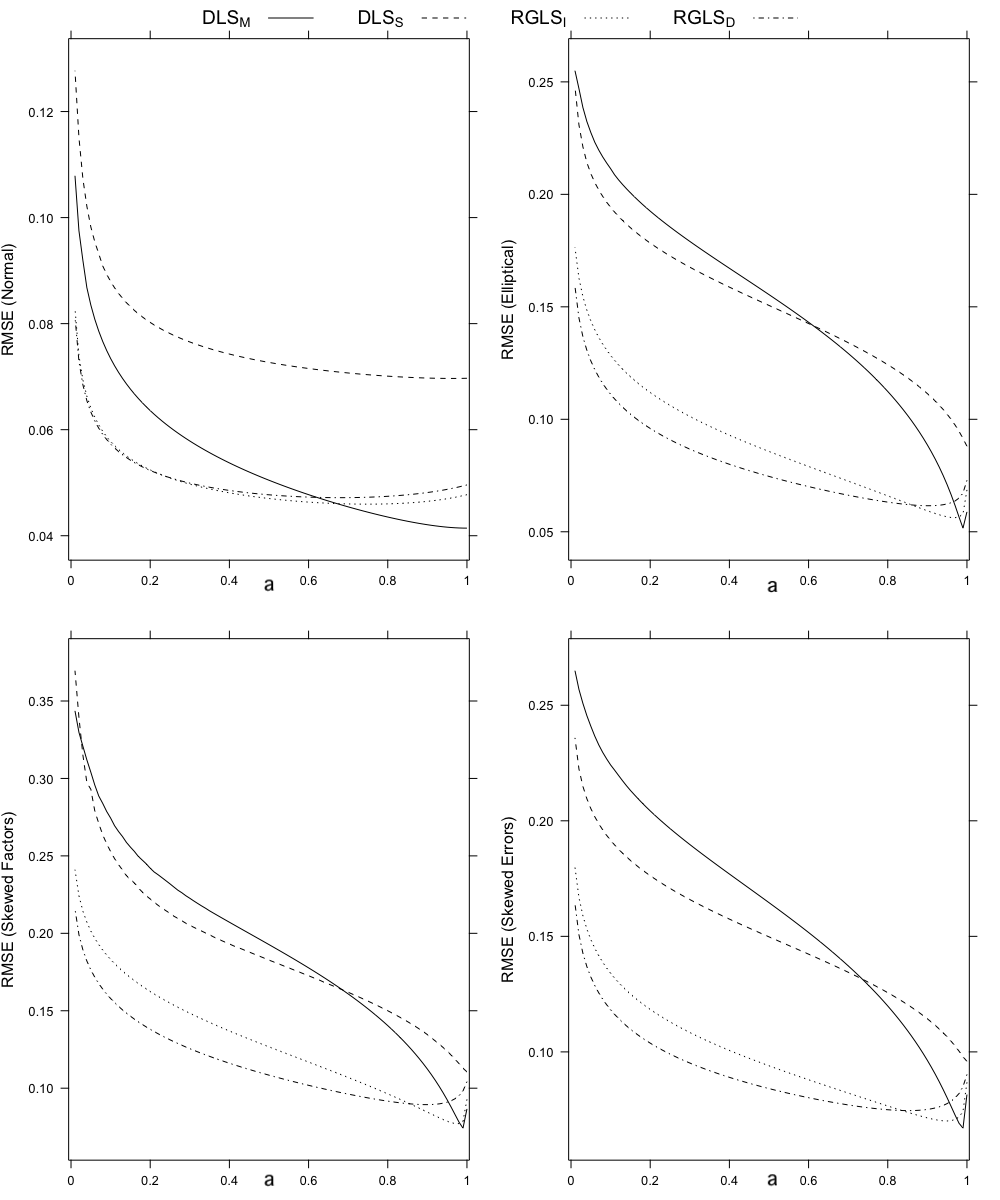}
\end{figure}

In terms of convergence rates, when $a$ was large, $DLS_{M}$ and
$DLS_{S}$ had no convergence issues. With a smaller $a$, $DLS_{M}$
and $DLS_{S}$ relied more on the ADF estimator. The ADF estimator
had convergence issues when the sample size $N$ was small relative
to the number of variables $p$, therefore $DLS_{M}$ and $DLS_{S}$
with small $a$ values could have low convergence rates. For example,
when $N=60$, $p=15$, $m=3$, and $a=0.1$, the convergence rates
ranged from 0.4 to 0.8 across all distributional conditions. With
a large enough $N$, the convergence rates were almost 1 even when
$a=0$. For example, when $N=200$, $p=15$, $m=3$, and $a=0$, the
convergence rates were above 0.95 across all distributional conditions.
With $a_{s}$, the convergence rates of $DLS_{M}$ were larger than
0.98 across all conditions, and the convergence rates of $DLS_{S}$
were larger than 0.91 (only when $p=15$, $m=3$, $N=40$ with skewed
factors, the convergence rate in $DLS_{S}$ was 0.735). 

The relative biases of the SE estimates were also influenced by $a$
in $DLS_{M}$ and $DLS_{S}$. Because factor loadings and covariances
between factors are usually the focus of research questions, we focus
on the relative biases of the SE estimates of factor loadings and
factor covariances. The SE estimates of residual variances were slightly
higher than those of factor loadings and factor covariances. Surprisingly,
when $p=5$ and $m=1$, $N\geq300$, and the data were normal, $a=1$
yielded the largest average relative biases for factor loadings and
factor covariances but smallest average relative biases for all parameters,
whereas $N<300$, $a=1$ yielded the smallest average relative biases
for factor loadings and factor covariances (see Figure 3 for $N=300$,
$p=5$ and $m=1$). When the data were normal with other models, $a=1$
yielded the smallest average relative biases for factor loadings and
factor covariances and smallest average relative biases for all parameters
(see Figure 4 for $N=300$, $p=30$ and $m=3$). When the data were
nonnormal, similar to RMSE, the $a$ which gave the minimal biases
was smaller with a simpler model (see Figures 3-4). Note that $a_{s}$
was not guaranteed to provide the smallest biases. For example, when
$N=300$, $p=5$, $m=1$, and the data were elliptical, the $a_{s}$
that provided the smallest average RMSE in $DLS_{M}$ was 0.27, but
the $a$ that provided the smallest average bias was 0.66.

\begin{figure}
{\footnotesize{}\caption{{\footnotesize{}Plot of average relative biases of the SE estimates
of factor loadings and factor covariances depending on $a$ when $N=300$,
$p=5$ and $m=1$}}
}{\footnotesize\par}

\begin{singlespace}
{\footnotesize{}Figure 3: Plot of average relative biases of the SE
estimates of factor loadings and factor covariances depending on $a$
when $N=300$, $p=5$ and $m=1$}{\footnotesize\par}
\end{singlespace}

\includegraphics[scale=0.5]{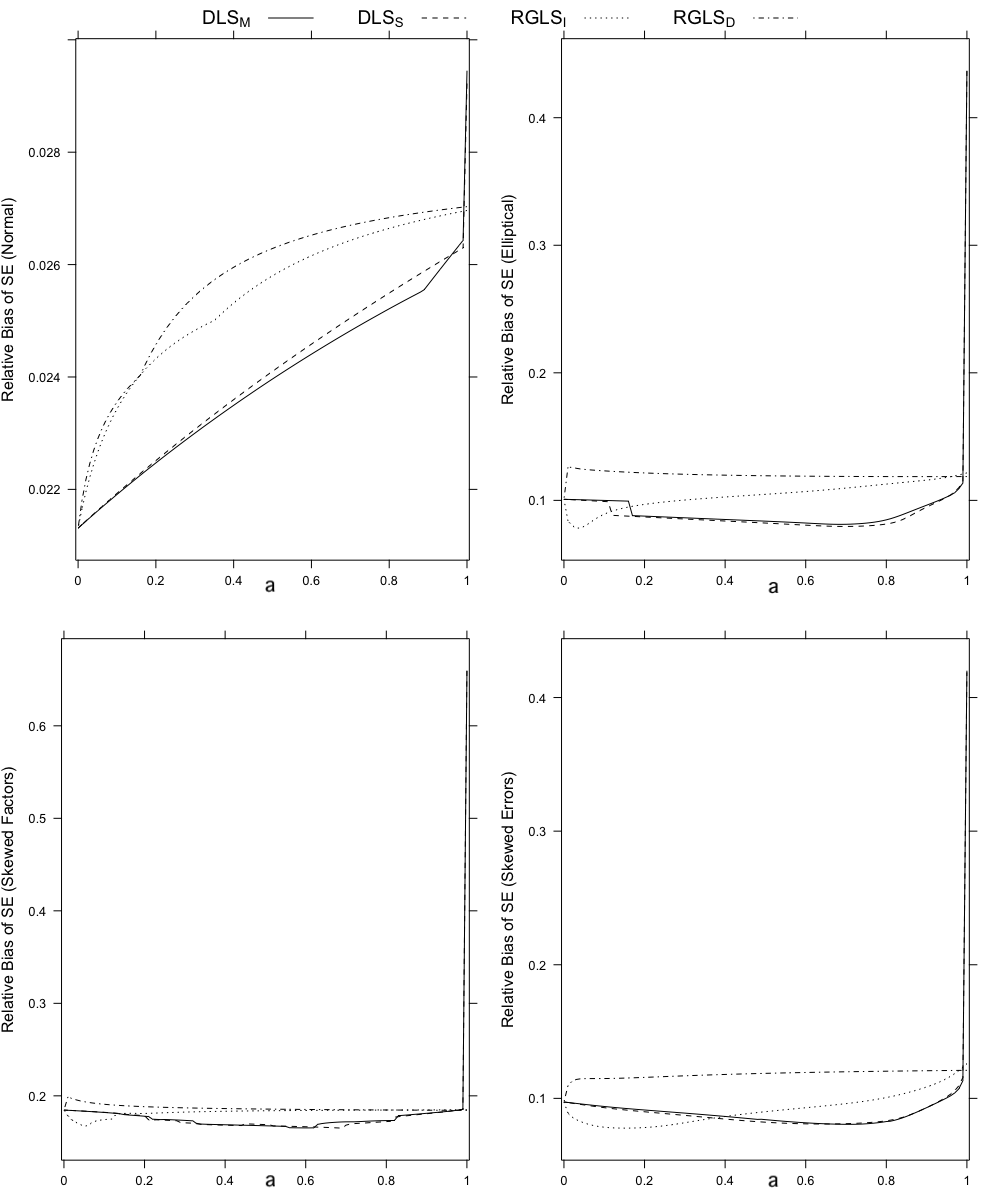}
\end{figure}
\begin{figure}
{\footnotesize{}\caption{{\footnotesize{}Plot of average relative biases of the SE estimates
of factor loadings and factor covariances depending on $a$ when $N=300$,
$p=30$ and $m=3$}}
}{\footnotesize\par}

\begin{singlespace}
{\footnotesize{}Figure 4: Plot of average relative biases of the SE
estimates of factor loadings and factor covariances depending on }$a$
when $N=300$, $p=30$ and $m=3$
\end{singlespace}

\includegraphics[scale=0.5]{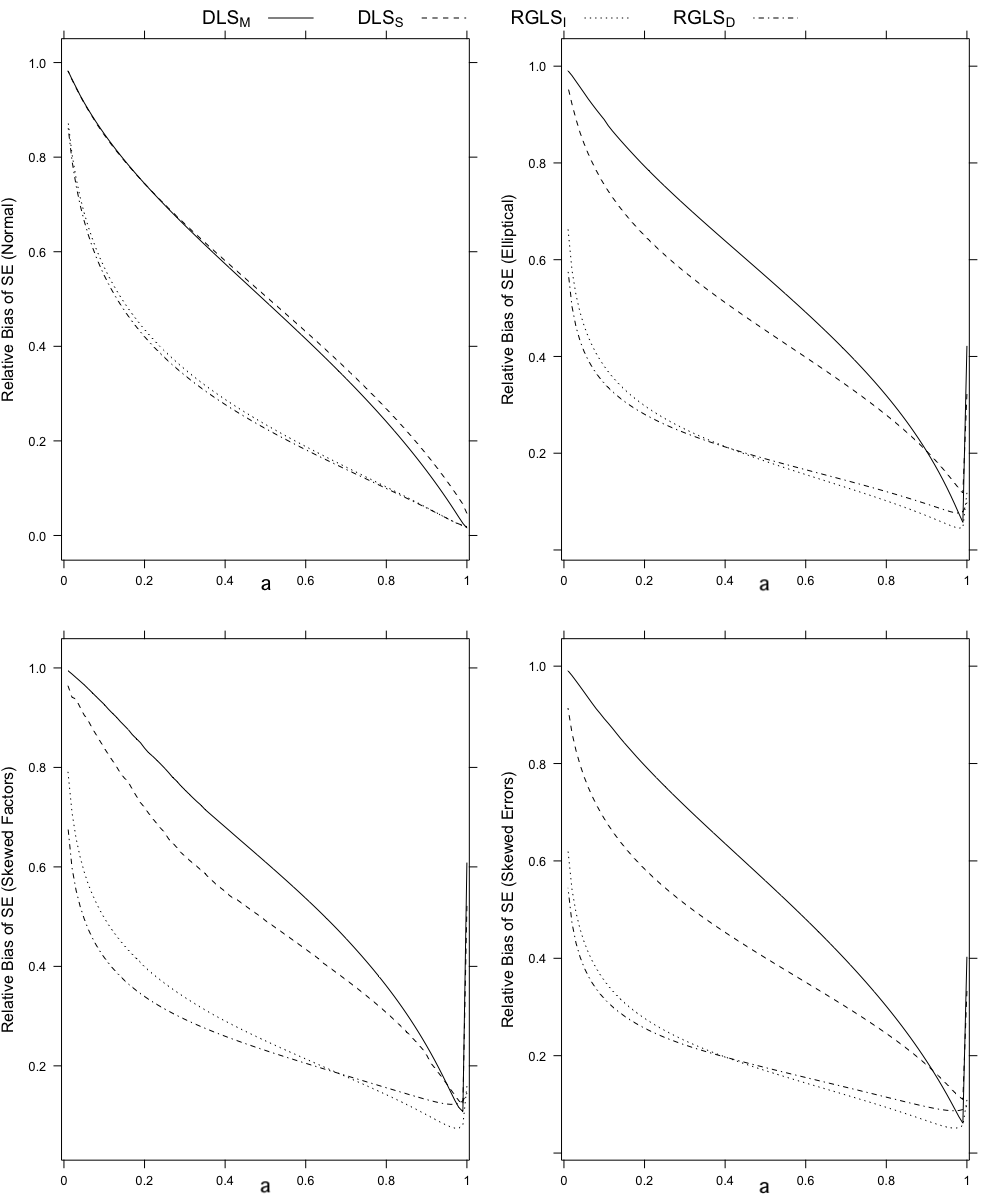}
\end{figure}

In addition, $a$ influenced Type I error rates from the standard
model fit statistic ($T$), the Satorra--Bentler test statistic ($T_{SB}$),
the mean and variance adjusted test statistic ($T_{MVA}$), the Jiang-Yuan
rank adjusted test statistic ($T_{JY}$). The influential pattern
depended on the sample size, the model complexity, and the distribution.

In general, $a$ influenced convergence rates (when $N$ was small),
RMSEs of parameter estimates, biases of standard error estimates,
and Type I error rates of model fit statistics in $DLS_{M}$ and $DLS_{S}$.
$a$'s influence depended on the sample size, the model complexity,
and the distribution. With the $a_{s}$ from the minimal RMSE, the
convergence rates of $DLS_{M}$ and $DLS_{S}$ were acceptable. When
we select $a_{s}$ based on the smallest RMSE, $a_{s}$ was not guaranteed
to provide the smallest biases of SE estimates.

\subsection{$RMSE$ across Methods}

We compared the 11 methods ($DLS_{M}$, $DLS_{S}$, $RGLS_{D}$, $RGLS_{I}$,
$LS$, $WLS$, $GLS_{S}$, $GLS_{M}$, $ML_{S}$, $ML_{O.M}$, and
$ML_{E.M}$) in terms of their efficiency and accuracy using the RMSE
across all sample sizes and models, and separately by distributional
conditions. Among $DLS_{M}$, $DLS_{S}$, $RGLS_{D}$, and $RGLS_{I}$,
the minimal RMSEs were selected given each method, and $a$ corresponding
to the smallest RMSE is referred to as $a_{s}$. We present the RMSEs
from all methods and all conditions in the supplemental material.
When the data were normal, we illustrate the RMSEs from the 11 methods
in Figure 5 with different $N$ and models. One overall pattern was
that when $N$ was larger, the RMSEs from the 11 methods became smaller.
The RMSEs from $DLS_{M}$ were the smallest among $DLS_{M}$, $DLS_{S}$,
$RGLS_{D}$ and $RGLS_{I}$, followed by $RGLS_{I}$ (see the upper
left panel of Figure 5). There were almost no differences of RMSEs
when the model was simple ($p=5$ and $m=1$). When the model became
more complex and $N$ was small, the sample covariance based DLS ($DLS_{S}$)
had a large RMSE which indicated inefficient and inaccurate estimation,
probably due to sample covariances not being stable with a small $N$.
Additionally, $RGLS_{I}$ performed better than $RGLS_{D}$, consistent
with the findings of Yuan and Chan (2016)\nocite{yuan2016structural}.
Among $LS$, $WLS$, $GLS_{S}$, and $GLS_{M}$, $GLS_{M}$ had the
smallest RMSEs (see the upper right panel of Figure 5). Similar to
$DLS_{M}$ and $DLS_{S}$, the model-implied covariance based GLS
($GLS_{M}$) outperformed the sample covariance based GLS ($GLS_{S}$).
The point estimates from $ML_{O.M}$, $ML_{E.M}$, and $ML_{S}$ were
the same and hence had the same RMSEs (see the lower left panel of
Figure 5). We select the methods that provided the smallest RMSEs
from the upper left panel, upper right panel, and lower left panel,
and plot them again in the lower right panel of Figure 5: $DLS_{M}$,
$RGLS_{I}$, $GLS_{M}$, and $ML$ ( $ML_{O.M}$, $ML_{E.M}$, and
$ML_{S}$ had the same point estimates and RMSEs). Although the RMSEs
in $DLS_{M}$ were also larger with a more complex model and a smaller
$N$, they were still smaller than the other methods (e.g., $RGLS_{I}$)
or equivalent to the normal theory based methods ($GLS_{M}$ and $ML$),
because with normal data, $a_{s}$ in $DLS_{M}$ was usually almost
1.

\begin{figure}
{\footnotesize{}\caption{{\footnotesize{}RMSEs from the 11 methods when data are normal}}
}{\footnotesize\par}

\begin{singlespace}
{\footnotesize{}Figure 5: RMSEs from the 11 methods when data are
normal}{\footnotesize\par}
\end{singlespace}

\includegraphics[scale=0.5]{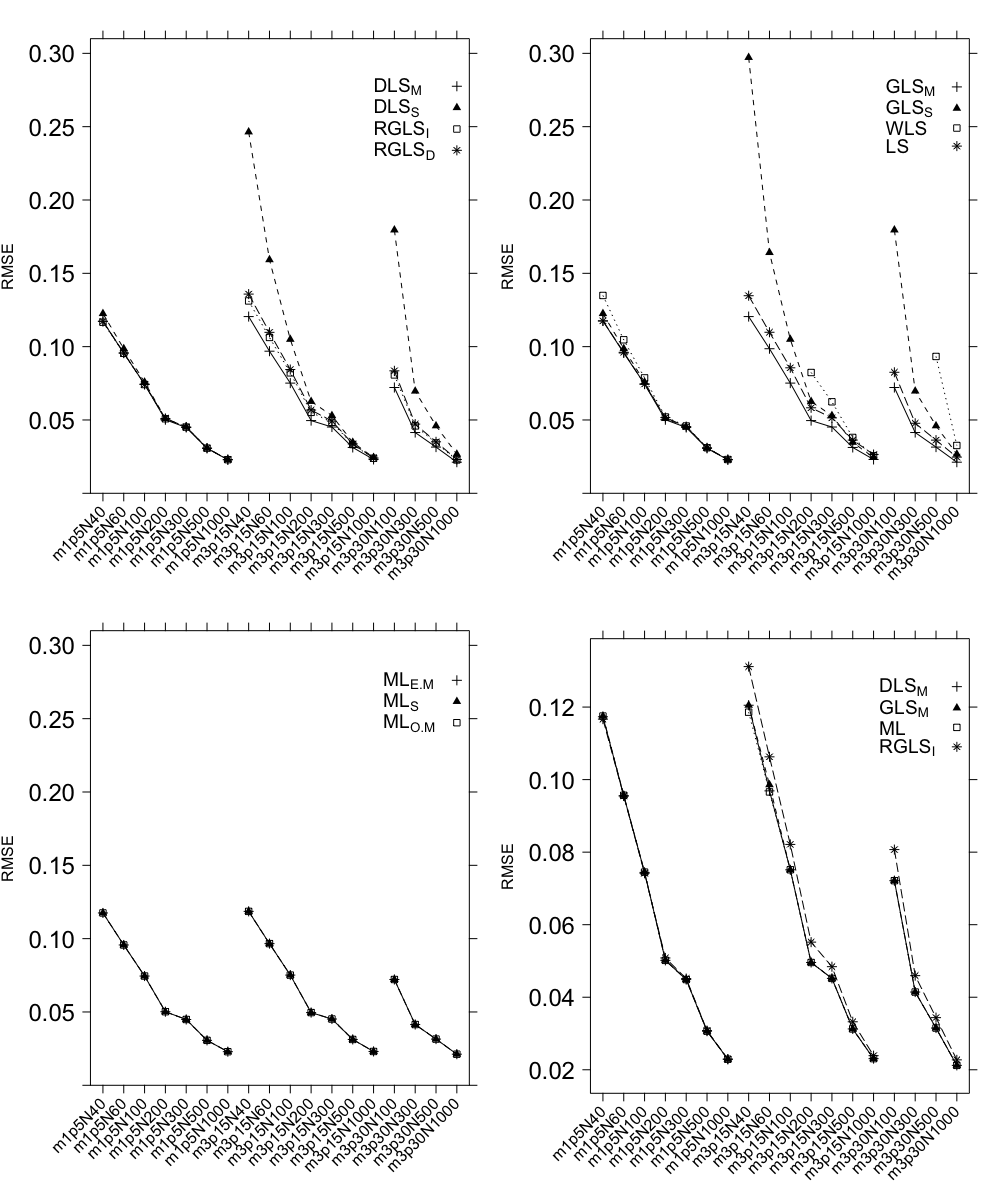}

Note: The $ML$ methods ($ML_{O.M}$, $ML_{E.M}$, and $ML_{S}$)
have the same parameter estimates but different standard errors, therefore
their RMSEs are the same. The y-axis of the lower right panel is different
from the other three panels.
\end{figure}

When the distributional condition was elliptical, the errors were
skewed, or the factors were skewed, the patterns of RMSEs from the
11 methods were similar across the distributional conditions, therefore
we present the elliptical condition as an example. Among $DLS_{M}$,
$DLS_{S}$, $RGLS_{D}$, and $RGLS_{I}$, $DLS_{M}$ had the smallest
RMSEs, followed by $RGLS_{I}$ (see the upper left panel of Figure
6). Among $LS$, $WLS$, $GLS_{S}$, and $GLS_{M}$, $GLS_{M}$ had
the smallest RMSEs (see the upper right panel of Figure 6). Due to
the instability of the sample covariances with small $N$s, $DLS_{S}$
and $GLS_{S}$ could yield large RMSEs. We select the methods that
provided the smallest RMSEs and plot them in the the lower right panel
of Figure 6 again: $DLS_{M}$, $RGLS_{I}$, $GLS_{M}$, and $ML$
(the types of information matrix and covariance matrix did not matter).
$DLS_{M}$ provided the smallest RMSEs, followed by $RGLS_{I}$. The
normal theory based methods, such as $GLS_{M}$ and $ML$, had larger
RMSEs compared to $DLS_{M}$. Such a difference of RMSEs was larger
when the error were skewed or the factors were skewed (e.g., 0.08).
This indicated that the ADF component in $DLS_{M}$ improved the efficiency
and accuracy of parameter estimation when data were nonnormal, while
the completely normal theory based methods provided somewhat less
accurate estimates due to assumption violations.

\begin{figure}
{\footnotesize{}\caption{{\footnotesize{}RMSEs from the 11 methods when data are elliptical}}
}{\footnotesize\par}

\begin{singlespace}
{\footnotesize{}Figure 6: RMSEs from the 11 methods when data are
elliptical}{\footnotesize\par}
\end{singlespace}

\includegraphics[scale=0.5]{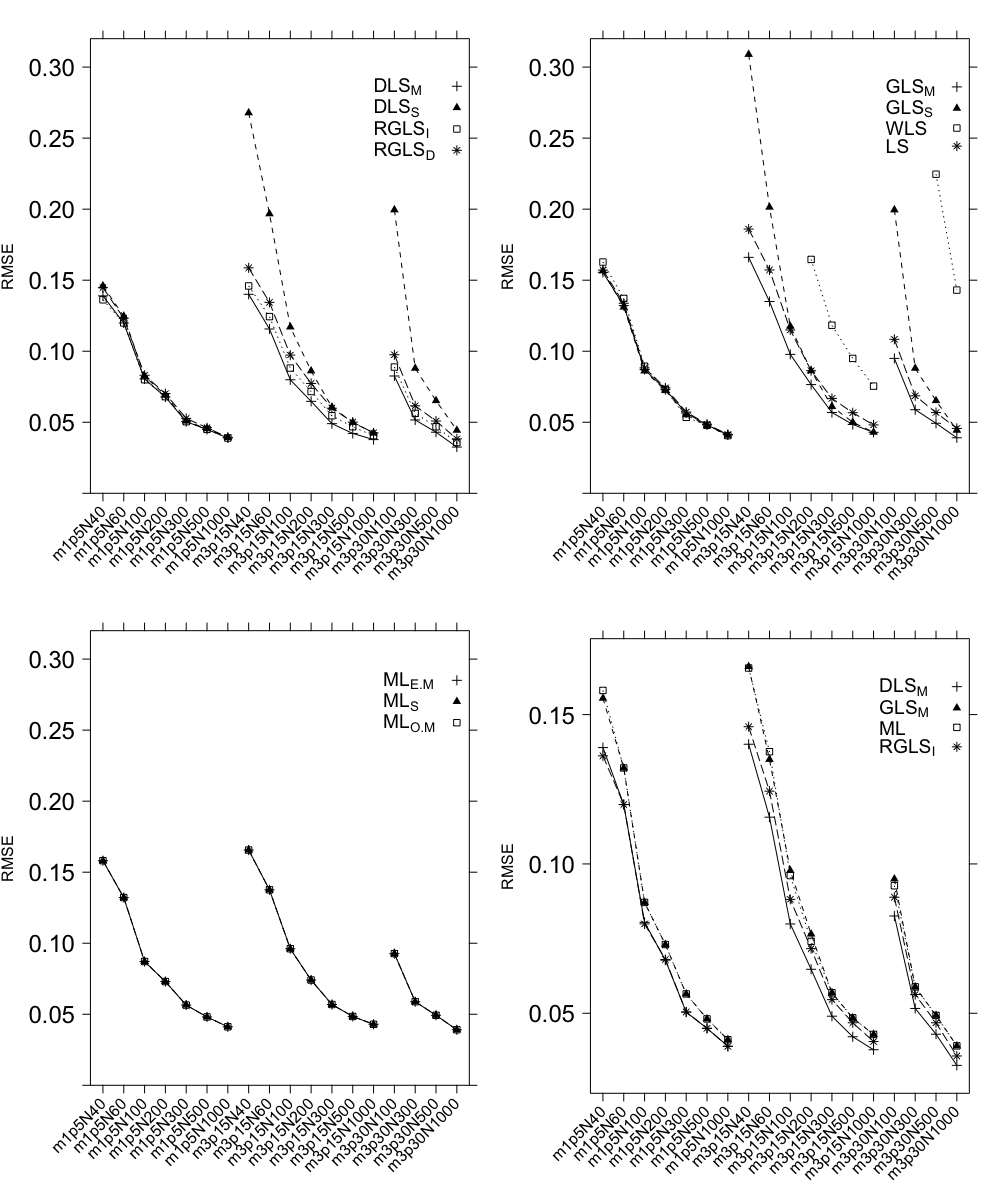}

Note: The $ML$ methods ($ML_{O.M}$, $ML_{E.M}$, and $ML_{S}$)
have the same parameter estimates but different standard errors, therefore
their RMSEs are the same. The y-axis of the lower right panel is different
from the other three panels.
\end{figure}

\subsection{Empirical SEs across Methods}

We focus on the empirical SEs of the estimates of factor loadings
and covariances between factors. The average empirical SEs of residual
variances were generally lower than those of factor loadings and factor
covariances. We present the empirical SEs from all methods and all
conditions in the supplemental material. When the data were normal,
the average empirical SEs of $DLS_{M}$ were the relatively smallest
and equivalent to the normal theory based methods ($GLS_{M}$ and
$ML$). When the data were nonnormal, $DLS_{M}$ and $RGLS_{I}$ had
the smallest average empirical SEs. Depending on the distributional
condition and $N$, $DLS_{M}$ or $RGLS_{I}$ could be smaller than
the other (see Figure 7 for an example for the skewed factor case).
Especially, when $N$ was small, $DLS_{M}$ could have smaller SEs
than $RGLS_{I}$. 

\begin{figure}
{\footnotesize{}\caption{{\footnotesize{}Empirical SEs of factor loadings and factor covariances
from the 11 methods when the factor are skewed}}
}{\footnotesize\par}

\begin{singlespace}
{\footnotesize{}Figure 7: Empirical SEs of the SE estimates of factor
loadings and factor covariances from the 11 methods when the factor
are skewed}{\footnotesize\par}
\end{singlespace}

\includegraphics[scale=0.5]{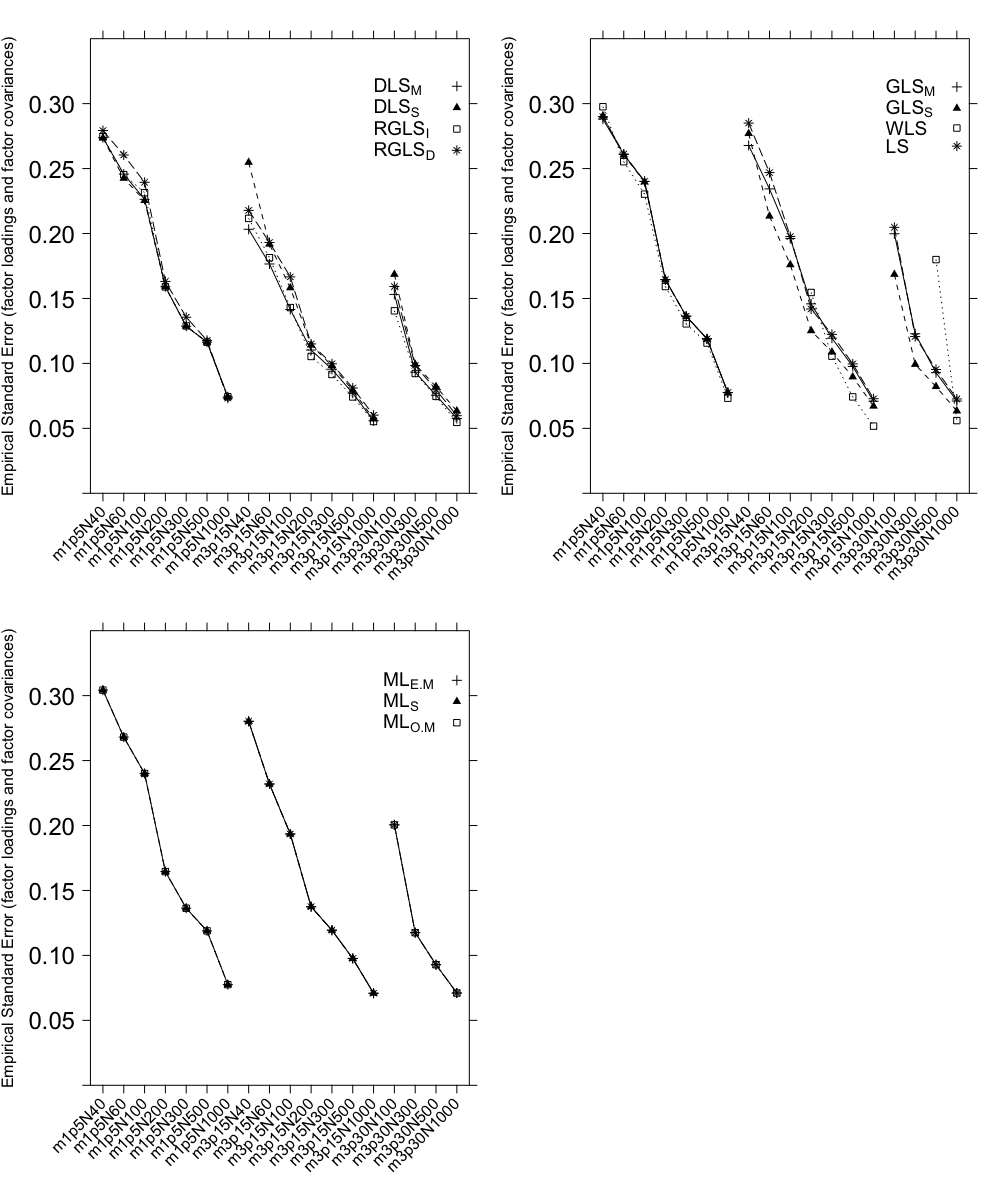}

Note: The $ML$ methods ($ML_{O.M}$, $ML_{E.M}$, and $ML_{S}$)
have the same parameter estimates but different standard errors, therefore
their empirical standard errors (SE) are the same. In this condition,
the methods providing the minimal empirical SEs are $DLS_{M}$, $DLS_{M}$
, $RGLS_{I}$, and $RGLS_{D}$. Because they are all in the upper
left panel , we do not create a lower right panel to summarize these
4 methods again.
\end{figure}

\subsection{Relative Biases of SE Estimates across Methods}

We compared the 11 methods in terms of their relative biases of the
SE estimates, averaging over factor loadings and factor covariances.
The presented $DLS_{M}$, $DLS_{S}$, $RGLS_{D}$, and $RGLS_{I}$
used the $a_{s}$ which provided the smallest RMSE. We present the
relative biases of SE estimates from all methods and all conditions
in the supplemental material. When the data were normal, the relative
biases of the SE estimates from the 11 methods with different $N$
and models are presented in Figure 8. Similar to the RMSEs, the overall
pattern was that when $N$ increased, the relative biases from the
11 methods generally became smaller. The biases of SE estimates from
$DLS_{M}$ were the smallest among $DLS_{M}$, $DLS_{S}$, $RGLS_{D}$
and $RGLS_{I}$, followed by $RGLS_{I}$ (see the upper left panel
of Figure 8). By our definition, the SE estimates of $DLS_{M}$ and
$DLS_{S}$ depended on $a_{s}$. With $a_{s}=1$, the SEs were standard
SE estimates; otherwise the SEs were sandwich SE estimates. Among
$LS$, $WLS$, $GLS_{S}$, and $GLS_{M}$, the sandwich SE estimates
of $LS$ had the smallest biases (see the upper right panel of Figure
8). Although $GLS_{S}$ and $GLS_{M}$ had the correctly specified
normal assumption, the sandwich SE estimates were calculated which
could increase the biases of SE estimates. The $ML$ sandwich SE estimates
based on the expected information and model implied covariance $ML_{E.M}$
provided the smallest biases among the $ML$ methods (see the lower
left panel of Figure 8). We select the methods that provided the smallest
biases of SE estimates in the upper left panel, upper right panel,
and lower left panel, and plot them in the lower right panel of Figure
8: $DLS_{M}$, $RGLS_{I}$, $LS$, and $ML_{E.M}$. The SE estimates
of $DLS_{M}$ had the smallest biases.

\begin{figure}
{\footnotesize{}\caption{{\footnotesize{}Average relative biases of the SE estimates of factor
loadings and factor covariances from the 11 methods when data are
normal}}
}{\footnotesize\par}

\begin{singlespace}
{\footnotesize{}Figure 8: Average relative biases of the SE estimates
of factor loadings and factor covariances from the 11 methods when
data are normal}{\footnotesize\par}
\end{singlespace}

\includegraphics[scale=0.5]{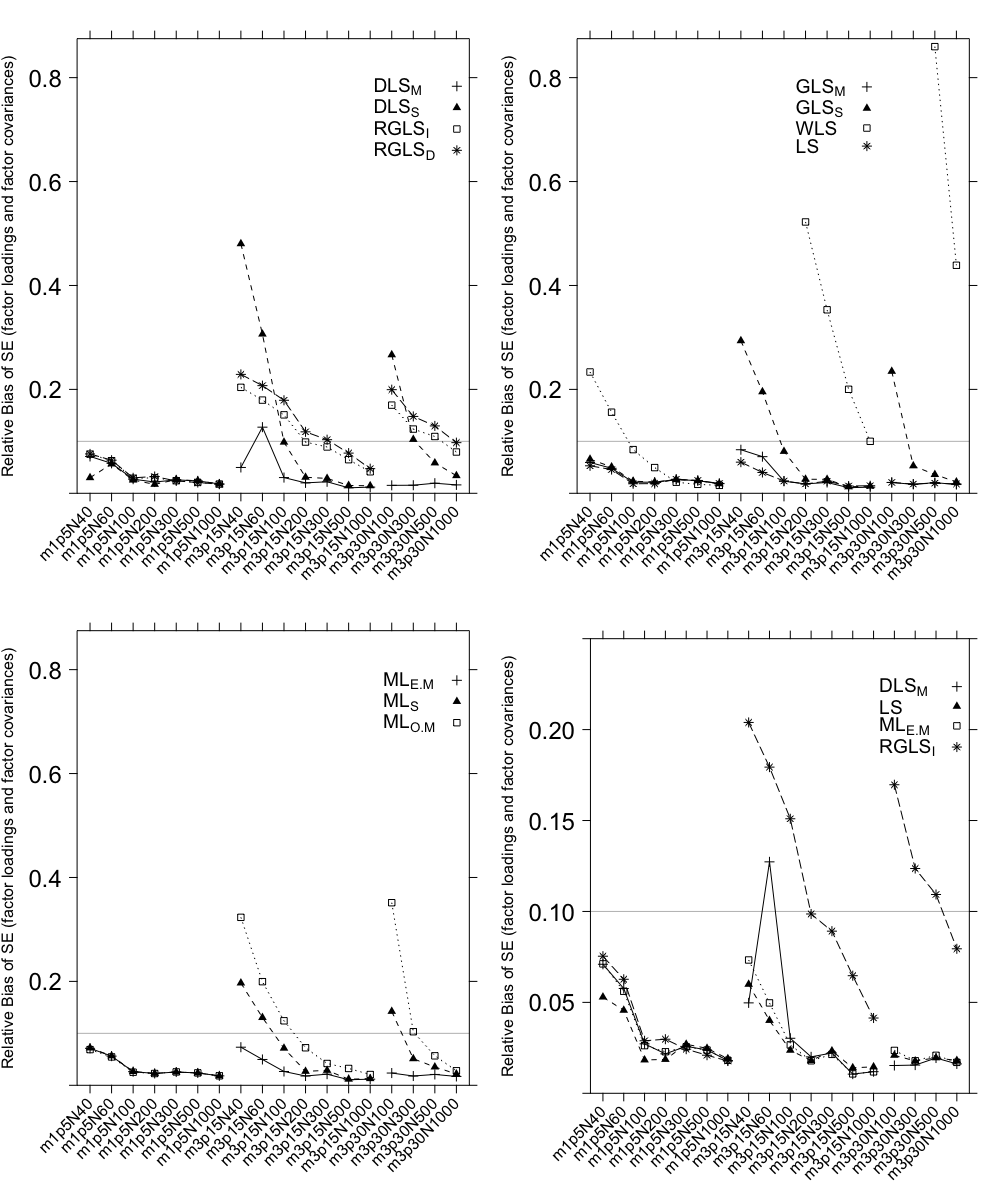}

Note: The y-axis of the lower right panel is different from the other
three panels. The grey line indicates a 10\% relative bias threshold.
\end{figure}

When the distributional condition was elliptical, errors were skewed,
and factors were skewed, the patterns of the relative biases of SE
estimate from the 11 methods were similar across the distributional
conditions, therefore we present the elliptical condition as an example.
Among $DLS_{M}$, $DLS_{S}$, $RGLS_{D}$, and $RGLS_{I}$, $RGLS_{I}$
had the smallest biases of SE estimates (see the upper left panel
of Figure 9). When $N$ was not extremely small, the SE estimates
of $DLS_{M}$ were similar to those of $RGLS_{I}$. Among $LS$, $WLS$,
$GLS_{S}$, and $GLS_{M}$, the sandwich SE estimates of $GLS_{M}$
and $LS$ had the similar biases (see the upper right panel of Figure
9). The $ML$ sandwich SE estimates based on the expected information
and sample covariance $ML_{E.M}$ provided the smallest biases among
the $ML$ methods (see the lower left panel of Figure 9). We select
the methods that provided the smallest biases of SE estimates and
plot them in the the lower right panel of Figure 9: $DLS_{M}$, $RGLS_{I}$,
$GLS_{M}$, and $ML_{E.M}$. $DLS_{M}$ or $RGLS_{I}$ generally had
the smallest biases of SE estimates. The SE estimates of $RGLS_{I}$
could be less biased compared to $DLS_{M}$ when $N$ was small.

\begin{figure}
{\footnotesize{}\caption{{\footnotesize{}Average relative biases of the SE estimates of factor
loadings and factor covariances from the 11 methods when data are
elliptical}}
}{\footnotesize\par}

\begin{singlespace}
{\footnotesize{}Figure 9: Average relative biases of the SE estimates
of factor loadings and factor covariances from the 11 methods when
data are elliptical}{\footnotesize\par}
\end{singlespace}

\includegraphics[scale=0.5]{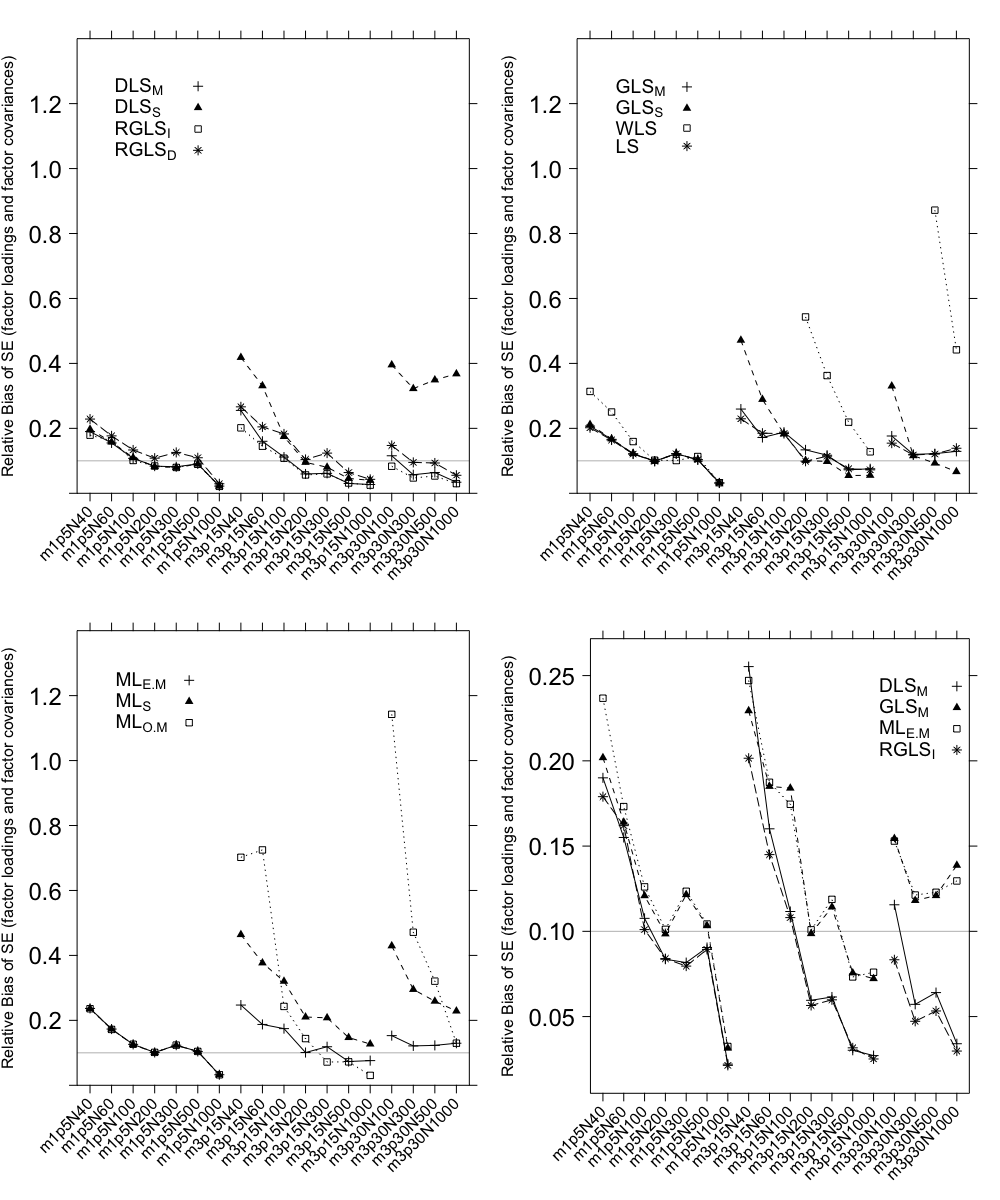}

Note: The y-axis of the lower right panel is different from the other
three panels. The grey line indicates a 10\% relative bias threshold.
\end{figure}

\subsection{Type I Error Rates across Methods}

We examined the Type I error rates of the standard model fit statistic
($T$), the Satorra--Bentler test statistic ($T_{SB}$), the mean
and variance adjusted test statistic ($T_{MVA}$), and the Jiang-Yuan
rank adjusted test statistic ($T_{JY}$) from the 11 methods from
the four distributional conditions. We consider a Type I error rate
between 0.025 and 0.075 as satisfactory \citep{bradley1978robustness}.
We present the Type I error rates from all methods and all conditions
in the supplemental material. $T$, $T_{SB}$, and $T_{MVA}$ yielded
either too small (e.g., 0) or too larger Type I error rates (e.g.,
1) with all 11 methods. We failed to find a method which uniformly
provided acceptable Type I error rates using $T$, $T_{SB}$, or $T_{MVA}$.
We examined the performance of $T_{JY}$ compared to both $\chi_{df}^{2}$
and $\chi_{rank\left(\mathbf{\hat{U}\hat{\Gamma}_{ADF}}\right)}^{2}$.
With referring to $\chi_{rank\left(\mathbf{\hat{U}\hat{\Gamma}_{ADF}}\right)}^{2}$,
the performance of $T_{JY}$ was better, whereas with referring to
$\chi_{df}^{2}$, the Type I error rates in most methods generally
were 0 when $N$ was small. Hence, we report the results of $T_{JY}$
with $\chi_{rank\left(\mathbf{\hat{U}\hat{\Gamma}_{ADF}}\right)}^{2}$.
There were 4 methods that performed relatively better than the others
using $T_{JY}$: $DLS_{M}$, $DLS_{S}$, $GLS_{M}$, and $GLS_{S}$.
Their Type I error rates of $T_{JY}$ across distributional conditions,
models, and $N$ are presented in Figure 10. When $N$ was too small,
$DLS_{M}$, $DLS_{S}$, $GLS_{M}$, and $GLS_{S}$ deviated from the
nominal level. As $N$ became larger, the Type I error rates were
more acceptable. $DLS_{M}$ almost always provided acceptable Type
I error rates unless $N$ was too small relative to the model complexity.
However, the ML methods with $T_{JY}$ ($\chi_{rank\left(\mathbf{\hat{U}\hat{\Gamma}_{ADF}}\right)}^{2}$)
could have too high Type I error rates.

\begin{figure}
{\footnotesize{}\caption{{\footnotesize{}Type I error rates of the Jiang-Yuan rank adjusted
test statistic ($T_{JY}$) from $DLS_{M}$, $DLS_{S}$, $GLS_{M}$,
and $GLS_{S}$}}
}{\footnotesize\par}

\begin{singlespace}
{\footnotesize{}Figure 10: Type I error rates of the Jiang-Yuan rank
adjusted test statistic ($T_{JY}$) from $DLS_{M}$, $DLS_{S}$, $GLS_{M}$,
and $GLS_{S}$}{\footnotesize\par}
\end{singlespace}

\includegraphics[scale=0.5]{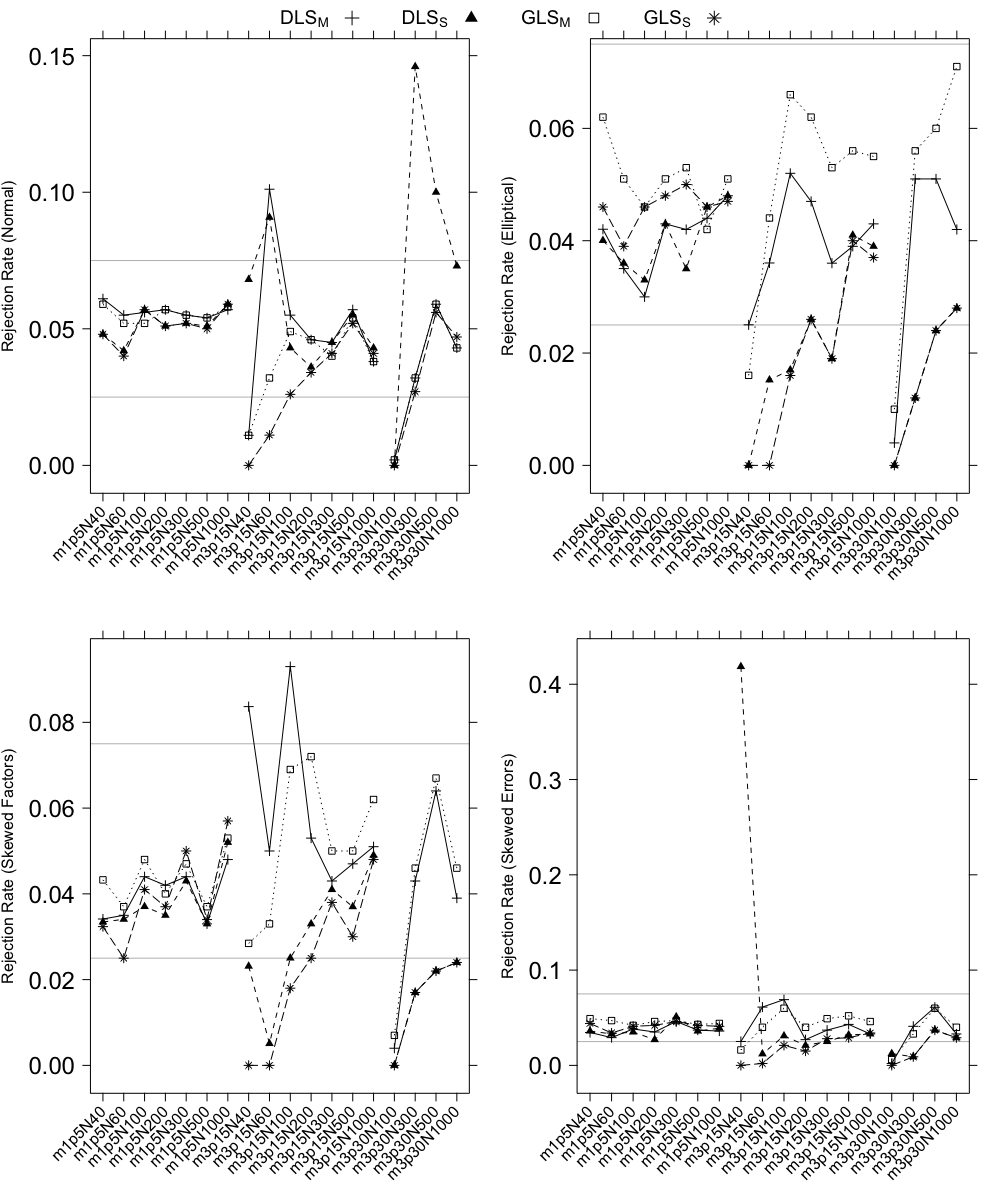}
\end{figure}

\subsection{Conclusions from the Simulation Study}

In summary, $DLS_{M}$ provided the smallest RMSEs regardless of the
distributions. When data were normal, $GLS_{M}$ and $ML$ provided
similar RMSEs as $DLS_{M}$; when data were nonnormal, $RGLS_{I}$
provided the second smallest RMSEs. In terms of the empirical SEs,
when data were normal, the empirical SE estimates of $DLS_{M}$, $GLS_{M}$,
and $ML$ were the smallest, and the SE estimates of $DLS_{M}$ had
the smallest biases. When data were nonnormal, the empirical SE estimates
of $DLS_{M}$ and $RGLS_{I}$ were the smallest, and the SE estimates
of $DLS_{M}$ and $RGLS_{I}$ had similar small biases while the SE
estimates of $RGLS_{I}$ could be less biased when $N$ was small.
Additionally, the Type I error rates of Jiang-Yuan rank adjusted test
statistic ($T_{JY}$) using $DLS_{M}$ were generally around the nominal
level (0.05). Overall, we recommend $DLS_{M}$ given its performance.

\subsection{Model Misspecification Simulation}

Since $DLS_{s}$ was inferior to $DLS_{M}$ in the simulation above,
we conducted a small-scale simulation study to explore whether $DLS_{s}$
outperformed $DLS_{M}$ when the model was misspecified. We considered
the case where $p=30$, $m=3$, $N$ varied as 100, 300, 500, and
1000, and the distributional condition was normal or elliptical. We
generated data the same as in the previous section, but we assumed
(1) the factor correlations were 0 or (2) all factor loadings were
equal. All the detailed results are presented in the supplemental
material. The patterns from the two types of misspecification did
not differ much. $DLS_{M}$ outperformed $DLS_{S}$ in terms of RMSEs,
the relative biases of SE estimates, and the empirical SEs. When data
were normal, as the sample size ($N$) increased, the difference between
$DLS_{S}$ and $DLS_{M}$ became smaller. When data were elliptical,
with a larger sample size, the difference between $DLS_{S}$ and $DLS_{M}$
in terms of RMSEs and empirical SEs became smaller but the difference
regarding the relative biases of SE estimates did not get smaller
(see Figure 11 for elliptical data and equal factor loading assumption
as an example). We found that $T_{JY}$ with $DLS_{M}$ generally
indicated a poor model fit across all sample sizes. It indicated that
consistent with the Type I error rate simulation results, $T_{JY}$
is an appropriate test statistic for $DLS_{M}$ for model fit evaluation.

\begin{figure}
{\footnotesize{}\caption{{\footnotesize{}RMSEs, relative biases of the SE estimates, and empirical
SEs for elliptical data and equal factor loading assumption }}
}{\footnotesize\par}

\begin{singlespace}
{\footnotesize{}Figure 11: RMSEs, relative biases of the SE estimates,
and empirical SEs for elliptical data and equal factor loading assumption}{\footnotesize\par}
\end{singlespace}

\includegraphics[scale=0.4]{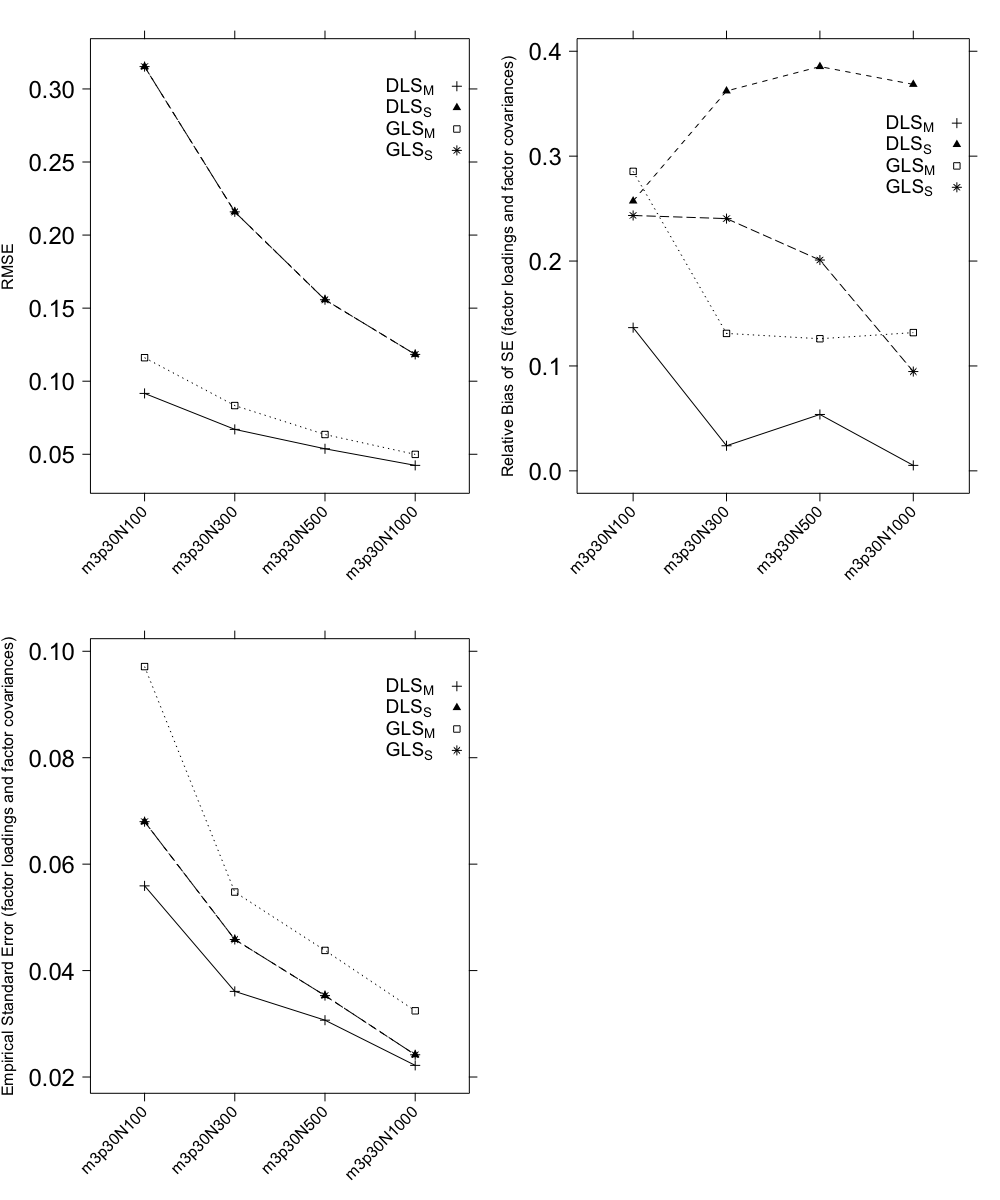}
\end{figure}

\section{Real Data Example}

In real data analyses, $a_{s}$ is unknown and needs to be estimated.
In this section, we illustrate how to apply the proposed distributionally-weighted
least squares estimation using a bootstrap procedure. We considered
$DLS_{M}$ and $RGLS_{I}$ that performed relatively well in the simulation
and two normal theory based methods, $ML_{S}$ and $GLS_{M}$. We
used a public dataset which is available in the R package, \textit{lavaan
(version 0.6-5)} \citep{rosseel2012lavaan}. The original dataset
from Holzinger and Swineford \citeyearpar{holzinger1939study} has
mental ability test scores of 26 tests for the 7th and 8th grade children
from two different schools (Pasteur and Grant-White). We focused on
a subset of 9 variables and 145 children from the Grant-White school
only. This subset is widely used in the SEM literature (e.g., \citealp{joreskog1969general,yuan2016structural}).
There are three dimensions/factors: spatial ability, verbal ability,
and ability related to speed. The 9th variable is the speeded discrimination
of straight and curved capitals. This variable measures both a spatial
ability and an ability related to speed, therefore it has loadings
on both factors. The factor model is Equation (\ref{eq:covariance equation})
with $\mathbf{\Psi}=diag\left(\psi_{11},\psi_{22},\psi_{33},\psi_{44},\psi_{55},\psi_{66},\psi_{77},\psi_{88},\psi_{99}\right)$,
\[
\mathbf{\Phi}=\left(\begin{array}{ccc}
1 & \phi_{12} & \phi_{13}\\
\phi_{12} & 1 & \phi_{23}\\
\phi_{13} & \phi_{23} & 1
\end{array}\right),
\]

\[
\mathbf{\Lambda}=\left(\begin{array}{ccccccccc}
\lambda{}_{11} & \lambda{}_{12} & \lambda{}_{13} & 0 & 0 & 0 & 0 & 0 & \lambda{}_{19}\\
0 & 0 & 0 & \lambda{}_{24} & \lambda{}_{25} & \lambda{}_{26} & 0 & 0 & 0\\
0 & 0 & 0 & 0 & 0 & 0 & \lambda{}_{37} & \lambda{}_{38} & \lambda{}_{39}
\end{array}\right).
\]

Following Yuan and Chan (2016)\nocite{yuan2016structural}, we conducted
a bootstrap study to evaluate the empirical RMSE using the estimated
parameters from the bootstrap samples. Let $\boldsymbol{x_{i}}$ be
a $9\times1$ vector of test scores for individual $i$, $\mathbf{S_{x}}$
be the sample covariance matrix for $\boldsymbol{x_{i}}$, and $\hat{\mathbf{\Sigma}}=\Sigma\left(\hat{\boldsymbol{\theta}}_{ML}\right)$
be the model-implied covariance matrix based on the ML estimates using
the raw data. First, we adopted the Bollen-Stine transformation for
bootstrapping \citep{bollen1992bootstrapping},
\begin{equation}
\boldsymbol{x_{i}^{(0)}}=\mathbf{\hat{\mathbf{\Sigma}}^{1/2}\mathbf{S_{x}^{-1/2}}}\boldsymbol{x_{i}.}\label{eq:transfer}
\end{equation}
Equation (\ref{eq:transfer}) is to create a new sample covariance
matrix. The sample covariance of $\boldsymbol{x_{i}}$ is $\mathbf{S_{x}}$,
while the sample covariance of $\boldsymbol{x}_{i}^{(0)}$ is $\hat{\mathbf{\mathbf{\Sigma}}}$.
After transforming, the null hypothesis (the factor model above) is
true and $\hat{\boldsymbol{\theta}}_{ML}$ gives the true population
parameters for $\boldsymbol{x}_{i}^{(0)}$. Second, we drew with replacements
of $\boldsymbol{x}_{i}^{(0)}$ to construct 1000 bootstrap samples.
We varied the tuning parameter $a$ from 0 to 1 with an equal interval
of .01 (i.e., 0, 0.01,..., 0.99, 1) and applied $DLS_{M}$ and $RGLS_{I}$
(with each $a$ value) to each bootstrap sample. We applied the same
RMSE equation as Equation (\ref{eq:rmse}) where $\hat{\boldsymbol{\theta}}_{ML}$
from the raw data $\boldsymbol{x_{i}}$ ($i=1,...,145$) was treated
as $\boldsymbol{\theta}$ to calculate an empirical RMSE. Under the
null hypothesis, the empirical RMSE is a consistent estimator of the
true RMSE \citep{yuan2016structural}.

We plot the empirical average RMSE of all the parameter estimates
for $DLS_{M}$ and $RGLS_{I}$ along with $a$ in Figure 12. $a_{s}$
was 0.75 in $DLS_{M}$ with the smallest RMSE at 0.094, and $a_{s}$
was 0.36 in $RGLS_{I}$ with the smallest RMSE at 0.097. Consistent
with the simulation results, $DLS_{M}$ yielded a smaller RMSE compared
to $RGLS_{I}$. Yuan and Chan (2016)\nocite{yuan2016structural} reported
$a_{s}$ at 0.35 but they considered $a$ from 0 to 1 with an equal
interval of .05. The difference between our replication of $RGLS_{I}$
and the result in Yuan and Chan (2016) is trivial. We conducted a
sensitivity test with respect to the number of bootstrap samples.
Besides 1000 samples, we also estimated $a_{s}$ using 500, 2000,
and 5000 samples. The estimated $a_{s}$ was always 0.36 in $RGLS_{I}$
and the estimated $a_{s}$ was 0.74 or 0.75 in $DLS_{M}$, which supported
the validation of the selected $a_{s}$ value.

\begin{figure}
{\footnotesize{}\caption{{\footnotesize{}Plot of root mean square error (RMSE) of model parameter
estimates depending on $a$ in the real data example}}
}{\footnotesize\par}

\begin{singlespace}
{\footnotesize{}Figure 12: Plot of root mean square error (RMSE) of
model parameter estimates depending on $a$ in the real data example}{\footnotesize\par}
\end{singlespace}

\includegraphics[scale=0.4]{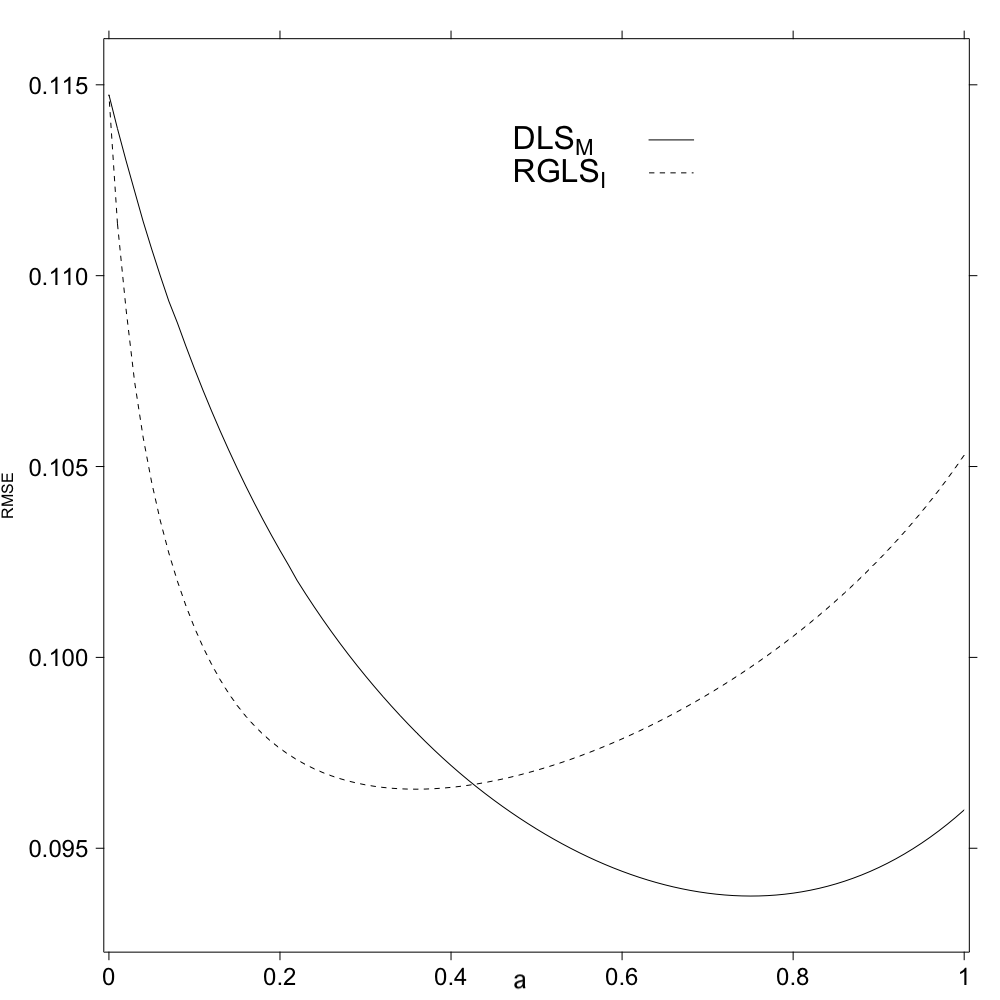}
\end{figure}

We applied $DLS_{M}\left(a_{s}=0.75\right)$, $RGLS_{I}\left(a_{s}=0.36\right)$,
$ML_{E.M}$, and $GLS_{M}$ to the raw data. The Jiang-Yuan rank adjusted
test statistic ($T_{JY}$) with $\chi_{rank\left(\mathbf{\hat{U}\hat{\Gamma}_{ADF}}\right)}^{2}$
for $DLS_{M}\left(a_{s}=0.75\right)$, $RGLS_{I}\left(a_{s}=0.36\right)$,
$ML_{E.M}$ and $GLS_{M}$ were 28.631 ($p=0.193$), 31.759 ($p=0.105$),
27.833 ($p=0.222$), and 27.255 ($p=0.245$), respectively. $DLS_{M}$,
$RGLS_{I}$, $ML_{E.M}$, and $GLS_{M}$ were shown to have good model
fits. The parameter estimates, the SE estimates, and the z scores
are in Table 3. There was little difference between the results of
the two normal theory based methods, $ML_{E.M}$ and $GLS_{M}$. $DLS_{M}\left(a_{s}=0.75\right)$
had similar results as $ML_{E.M}$ and $GLS_{M}$ except that $\psi_{88}$
is not statistically significantly different from 0. $RGLS_{I}\left(a_{s}=0.36\right)$
reached the similar significance conclusion as $DLS_{M}\left(a_{s}=0.75\right)$
although $\hat{\boldsymbol{\theta}}$s and SEs were slightly different.

\begin{table}
\caption{Real Data Example}

\resizebox{\textwidth}{!}{
\renewcommand{\arraystretch}{1.2}

{\huge{}}%
\begin{tabular}{c|ccc|ccc|ccc|ccc}
\hline 
\multicolumn{1}{c}{} & \multicolumn{3}{c|}{ $ML_{E.M}$ } & \multicolumn{3}{c|}{ $GLS_{M}$ } & \multicolumn{3}{c|}{ $DLS_{M}\left(a_{s}=0.75\right)$} & \multicolumn{3}{c}{$RGLS_{I}\left(a_{s}=0.36\right)$}\tabularnewline
\cline{2-13} \cline{3-13} \cline{4-13} \cline{5-13} \cline{6-13} \cline{7-13} \cline{8-13} \cline{9-13} \cline{10-13} \cline{11-13} \cline{12-13} \cline{13-13} 
\multicolumn{1}{c}{} & $\hat{\boldsymbol{\theta}}$  & SE & z & $\hat{\boldsymbol{\theta}}$  & SE & z & $\hat{\boldsymbol{\theta}}$  & SE & z & $\hat{\boldsymbol{\theta}}$  & SE & z\tabularnewline
\hline 
$\lambda{}_{11}$ & 0.817 & 0.099 & 8.263 & 0.817 & 0.109 & 7.482 & 0.811 & 0.1 & 8.148 & 0.797 & 0.098 & 8.145\tabularnewline
$\lambda{}_{12}$ & 0.541 & 0.1 & 5.426 & 0.541 & 0.094 & 5.778 & 0.531 & 0.088 & 6.059 & 0.516 & 0.085 & 6.073\tabularnewline
$\lambda{}_{13}$ & 0.686 & 0.09 & 7.642 & 0.686 & 0.088 & 7.808 & 0.685 & 0.083 & 8.235 & 0.699 & 0.083 & 8.392\tabularnewline
$\lambda{}_{19}$ & 0.458 & 0.089 & 5.126 & 0.458 & 0.103 & 4.444 & 0.503 & 0.095 & 5.278 & 0.507 & 0.088 & 5.778\tabularnewline
$\lambda_{24}$ & 0.972 & 0.078 & 12.383 & 0.972 & 0.084 & 11.596 & 0.958 & 0.08 & 11.988 & 0.948 & 0.078 & 12.123\tabularnewline
$\lambda_{25}$ & 0.96 & 0.083 & 11.631 & 0.96 & 0.083 & 11.541 & 0.956 & 0.079 & 12.092 & 0.94 & 0.078 & 12.031\tabularnewline
$\lambda_{26}$ & 0.934 & 0.081 & 11.553 & 0.934 & 0.083 & 11.272 & 0.928 & 0.08 & 11.591 & 0.916 & 0.083 & 11.092\tabularnewline
$\lambda_{37}$ & 0.705 & 0.09 & 7.853 & 0.705 & 0.084 & 8.383 & 0.669 & 0.078 & 8.549 & 0.639 & 0.081 & 7.917\tabularnewline
$\lambda_{38}$ & 0.897 & 0.093 & 9.598 & 0.897 & 0.098 & 9.138 & 0.919 & 0.094 & 9.77 & 0.925 & 0.097 & 9.506\tabularnewline
$\lambda_{39}$ & 0.45 & 0.09 & 5.032 & 0.451 & 0.094 & 4.812 & 0.401 & 0.088 & 4.54 & 0.405 & 0.082 & 4.91\tabularnewline
$\psi_{11}$ & 0.652 & 0.117 & 5.551 & 0.652 & 0.16 & 4.078 & 0.615 & 0.137 & 4.494 & 0.598 & 0.129 & 4.619\tabularnewline
$\psi_{22}$ & 0.933 & 0.122 & 7.634 & 0.933 & 0.142 & 6.586 & 0.933 & 0.131 & 7.111 & 0.929 & 0.124 & 7.478\tabularnewline
$\psi_{33}$ & 0.603 & 0.096 & 6.254 & 0.603 & 0.096 & 6.251 & 0.601 & 0.092 & 6.565 & 0.568 & 0.094 & 6.046\tabularnewline
$\psi_{44}$ & 0.313 & 0.065 & 6.4 & 0.313 & 0.066 & 4.735 & 0.318 & 0.063 & 5.05 & 0.31 & 0.067 & 4.62\tabularnewline
$\psi_{55}$ & 0.419 & 0.072 & 4.847 & 0.419 & 0.072 & 5.83 & 0.395 & 0.063 & 6.247 & 0.387 & 0.066 & 5.9\tabularnewline
$\psi_{66}$ & 0.408 & 0.069 & 5.824 & 0.408 & 0.077 & 5.329 & 0.399 & 0.073 & 5.476 & 0.402 & 0.075 & 5.37\tabularnewline
$\psi_{77}$ & 0.565 & 0.096 & 5.913 & 0.565 & 0.083 & 6.778 & 0.575 & 0.078 & 7.343 & 0.577 & 0.08 & 7.245\tabularnewline
$\psi_{88}$ & 0.289 & 0.118 & 5.865 & 0.289 & 0.13 & 2.224 & 0.179 & 0.121 & 1.477 & 0.104 & 0.134 & 0.773\tabularnewline
$\psi_{99}$ & 0.476 & 0.065 & 2.448 & 0.476 & 0.076 & 6.274 & 0.471 & 0.073 & 6.485 & 0.473 & 0.071 & 6.683\tabularnewline
$\phi_{12}$ & 0.554 & 0.081 & 6.86 & 0.554 & 0.092 & 6.026 & 0.583 & 0.085 & 6.844 & 0.603 & 0.081 & 7.477\tabularnewline
$\phi_{13}$ & 0.393 & 0.103 & 3.804 & 0.393 & 0.113 & 3.488 & 0.411 & 0.105 & 3.923 & 0.362 & 0.105 & 3.465\tabularnewline
$\phi_{23}$ & 0.239 & 0.095 & 2.511 & 0.239 & 0.118 & 2.023 & 0.233 & 0.115 & 2.02 & 0.244 & 0.113 & 2.16\tabularnewline
\hline 
\end{tabular}{\huge\par}

}
\end{table}

\section{Conclusion}

In real data analysis, data are unlikely to be exactly normally distributed.
If we ignore the non-normality reality, the parameter estimates, standard
error estimates, and model fit statistics from normal theory based
methods (e.g., ML and GLS) are unreliable. Even with the help of robust
statistics, the normal theory based methods' performances are not
adequate with finite sample sizes. On the other hand, the asymptotically
distribution free (ADF) estimator (i.e., WLS) does not rely on any
distribution assumption but cannot demonstrate its efficiency advantage
with small and modest sample sizes. We propose a distributionally-weighted
least squares (DLS) estimator, and expect that it can perform better
than the existing generalized least squares, because it combines normal
theory based and ADF based generalized least squares estimation. And
there are sample covariance based DLS ($DLS_{S}$) and model-implied
covariance based DLS ($DLS_{M}$).

Computer simulation results suggest that $DLS_{M}$ provides relatively
accurate and efficient estimates. Compared to ML estimators ($ML_{S}$,
$ML_{O.M}$, and $ML_{E.M}$), LS, WLS, $RGLS_{D}$, $RGLS_{I}$,
$GLS_{M}$, and $GLS_{S}$, $DLS_{M}$ provided the smallest RMSEs
regardless of the distributions. With normal data, $DLS_{M}$ had
the relatively smaller empirical SE estimates and smallest biases
of SE estimates, which were similar to those of $ML_{E.M}$ and $LS$;
with nonnormal data, $DLS_{M}$ and $RGLS_{I}$ had relatively smallest
empirical SE estimates and smallest biases, while $RGLS_{I}$'s SE
estimates could be smaller and less biased when $N$ was small. When
DLS coupled with the Jiang-Yuan rank adjusted test statistic ($T_{JY}$),
$DLS_{M}$ generally provided Type I error rates close to the nominal
level unless $N$ was too small relative to the model complexity.
However, $DLS_{S}$ did not perform well in terms of RMSEs, biases
of SE estimates, and model fit Type I error rates. Overall, $DLS_{M}$
is competitive with the existing methods in different aspects. The
simulation findings echo our anticipations at the beginning of the
paper: (1) $DLS_{M}$ yields more accurate and efficient estimates
than those from the  ADF estimator (WLS), (2) $DLS_{M}$ boosts convergence
rate compared to WLS, and (3) using data information while holding
the normality assumption to some degree enhances $DLS_{M}$'s performance.
But compared to $RGLS_{I}$, $DLS_{M}$ was shown to be more sensitive
to the selection of $a$ (Figures 2 and 4), which can be a reason
for one to use $RGLS_{I}$ instead of $DLS_{M}$. 

In the simulation, we explored the performance of $T_{JY}$ by referring
to $\chi_{df}^{2}$ and $\chi_{rank\left(\mathbf{\hat{U}\hat{\Gamma}_{ADF}}\right)}^{2}$.
We found when referring to $\chi_{df}^{2}$, the Type I error rates
in most methods generally were 0 when $N$ was small. When referring
to $\chi_{rank\left(\mathbf{\hat{U}\hat{\Gamma}_{ADF}}\right)}^{2}$,
multiple methods including $DLS_{M}$ provided acceptable Type I error
rates, whereas the ML methods could be too liberal. It should be recognized
that general methods for statistical model evaluation, such as the
Monte Carlo approach of Jalal and Bentler \citeyearpar{jalal2018using},
can also be adapted to evaluate DLS results.

A consequence of the availability of a method (i.e., $T_{JY}$) with
an acceptable Type I error rate is that the noncentral $\chi^{2}$
distribution is a good candidate to describe the behavior of DLS tests
under conditions of not-too-large misspecification. Hence methods
of describing model adequacy based on noncentrality-based fit indices
such as RMSEA (Steiger \& Lind, see \citealp{steiger2016notes}) and
CFI (Bentler, 1990)\nocite{bentler1990comparative} should be able
to be utilized. Illustrative recent research on these indices is Lai
\citeyearpar{lai2019simple,lai2020confidence}, Lai and Green \citeyearpar{lai2016problem},
Moshagen and Auerswald \citeyearpar{moshagen2018congruence}, and
Zhang and Savalei \citeyearpar{zhang2016bootstrapping,zhang2020examining}.
It also makes sense to consider new descriptive indices (e.g., \citealp{gomer2019new}),
or even traditional ones such as NFI and SRMR (e.g., \citealp{bentler1995eqs};
see also \citealp{maydeu2017assessing}) for use with DLS. 

The value of the tuning parameter $a$ determines the performance
of $DLS_{M}$. In the simulation, we selected the $a$ which yielded
the smallest RMSEs in $DLS_{M}$ and referred to it as $a_{s}$. We
were able to calculate RMSEs because we knew the population parameters
in the simulation. In practice, as we illustrated in the real data
example, we can use a bootstrap procedure to calculate an empirical
RMSEs to select the $a_{s}$. $a$ influenced convergence rates. With
a small $a$, $DLS_{M}$ could have nonconvergence when N was small
relative to the model complexity. But $a_{s}$ was not selected to
be small based on RMSEs. $a$ also influenced RMSEs, biases of SE
estimates, and model fit Type I error rates of $DLS_{M}$. The influence
depended on the distribution and the complexity of the model. 

In the simulation, we found that the selection of $a_{s}$ was based
on various factors such as the sample size, model complexity, and
distribution. In practice, we can estimate $a_{s}$ using the bootstrap
procedure by assuming the null hypothesis and the model implied covariance
are true. The selection of $a_{s}$ and the DLS inferences are based
on this assumption. In other words, the accuracy of estimated $a_{s}$
depends on how well the model fits the data. If the null hypothesis
is true, the bootstrapping results should be the same as our simulation
results. If the null hypothesis deviates from the true data generating
model, the estimated $a_{s}$ is not the best $a$ for the true model
but the best $a$ for the assumed model. As illustrated in the misspecification
simulation, with a misspecified model, $DLS_{M}$ with the estimated
$a_{s}$ generally performed better or was equivalent to the normal
theory based methods, $GLS_{M}$ and $GLS_{S}$, in terms of RMSE,
the biases of SE estimates, and the empirical SEs. We think it is
the best we can do given our model assumption. Additionally, when
analyzing real data, we suggest conducting a sensitivity analysis
as we did in the real data example. The estimated $a_{s}$ should
not change once the number of bootstrap samples is large enough. Another
option to select $a_{s}$ is to follow the work by \citet{jiang2018ridge},
\citet{yang2018ridge}, Yang and Yuan \citeyearpar{yang2019optimizing},
which constructs a mapping function between $a_{s}$ and all data/model
features. The selected $a_{s}$ should be more accurate than that
from the bootstrap procedure. But this approach requires an extensive
simulation to consider a variety of data/model features. 

Due to the scope and word limitation of this paper, we did not explore
the performance of the proposed method with missing data (e.g., multiple
imputation; \citealp{du2020model,enders2019model}). Future work could
look into investigating the performance of the proposed method with
missing data. 

In sum, our paper outlines a new distributionally-weighted least squares
estimator, $DLS_{M}$, which works well with both normal and nonnormal
data. $DLS_{M}$ can provide more accurate and efficient estimates
than classical methods, and the combination of $T_{JY}$ and $DLS_{M}$
provides acceptable Type I error rates.

\bibliographystyle{apalike}
\addcontentsline{toc}{section}{\refname}\bibliography{sem}

\end{document}